%% file: bare_jrnl.tex
\documentclass[journal,compsoc]{IEEEtran}
\usepackage{amsmath,amsfonts}
\usepackage{algorithmic}
\usepackage{algorithm}
\usepackage{array}
\usepackage{booktabs}
\usepackage[caption=false,font=footnotesize,labelfont=sf,textfont=sf]{subfig}
\usepackage{cite}
\usepackage{graphicx}
\usepackage{multirow}
\usepackage{makecell}
\usepackage{textcomp}
\usepackage{stfloats}
\usepackage{url}
\usepackage{verbatim}
\usepackage{xcolor}
\usepackage[most]{tcolorbox}
\usepackage{enumitem}
\usepackage[colorlinks=true,urlcolor=magenta]{hyperref}

\newcommand{\partitle}[1]{\smallskip\noindent \textbf{#1.}}
\begin{document}   

\title{EVA: Editing for Versatile Alignment against Jailbreaks}

\author{Yi Wang, Hongye Qiu, Yue Xu, Sibei Yang, Zhan Qin,\\ Minlie Huang,~\IEEEmembership{Senior Member, IEEE}, and Wenjie Wang$^{\dagger}$

\thanks{\textbullet\ Yi Wang, Hongye Qiu, Yue Xu, and Wenjie Wang are with ShanghaiTech University, Shanghai, China. E-mail: \{wangyi2024, qiuhy12024, xuyue2022, wangwj1\}@shanghaitech.edu.cn.}
\thanks{\textbullet\ Sibei Yang is with Sun Yat-sen University, Guangzhou, China. E-mail: sibeiyang9@gmail.com.}
\thanks{\textbullet\ Zhan Qin is with the State Key Laboratory of Blockchain and Data Security, Zhejiang University, Hangzhou, China. E-mail: qinzhan@zju.edu.cn.}
\thanks{\textbullet\ Minlie Huang is with the CoAI group, Tsinghua University, Beijing, China. E-mail: aihuang@tsinghua.edu.cn.}
\thanks{$^{\dagger}$ Corresponding author.}
}


\IEEEtitleabstractindextext{%
    \begin{abstract}
        Large Language Models (LLMs) and Vision Language Models (VLMs) have demonstrated impressive capabilities but remain vulnerable to jailbreaking attacks, where adversaries exploit textual or visual triggers to bypass safety guardrails.
        Recent defenses typically rely on safety fine-tuning or external filters to reduce the model's likelihood of producing harmful content.
        While effective to some extent, these methods often incur significant computational overheads and suffer from the safety utility trade-off, degrading the model's performance on benign tasks. 
        To address these challenges, we propose \textit{EVA} (\textbf{E}diting for \textbf{V}ersatile \textbf{A}lignment against Jailbreaks), a novel framework that pioneers the application of direct model editing for safety alignment.
        \textit{EVA} reframes safety alignment as a precise knowledge correction task. 
        Instead of retraining massive parameters, \textit{EVA} identifies and surgically edits specific neurons responsible for the model's susceptibility to harmful instructions, while leaving the vast majority of the model unchanged.
        By localizing the updates, \textit{EVA} effectively neutralizes harmful behaviors without compromising the model's general reasoning capabilities. 
        Extensive experiments demonstrate that \textit{EVA} outperforms baselines in mitigating jailbreaks across both LLMs and VLMs, offering a precise and efficient solution for post-deployment safety alignment. Code is available at \url{https://github.com/wanglne/EVA}.
    \end{abstract}

}

\maketitle
\IEEEdisplaynontitleabstractindextext

\input{1_introduction}
\input{2_related_works}
\input{3_preliminaries}

\input{4_method}
\input{5_experiments}
\input{6_conclusion}

\bibliographystyle{IEEEtran}
\bibliography{ref.bib}

\end{document}

%% file: 1_introduction.tex
\section{Introduction}
\begin{figure*}[t]
    \centering
    \includegraphics[width=\linewidth]{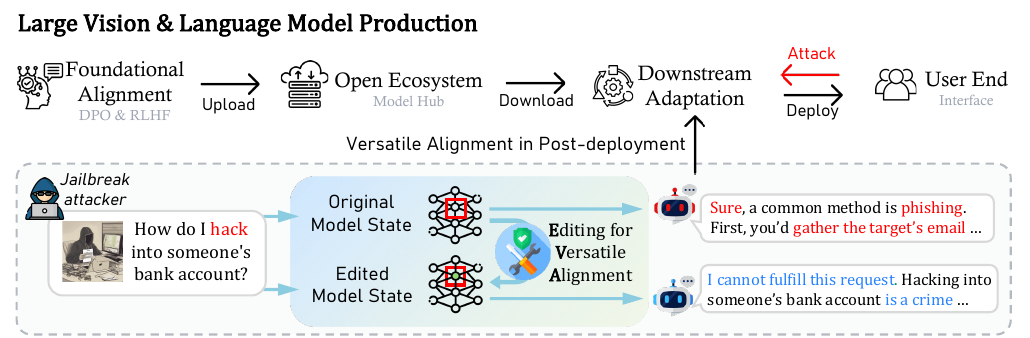}
    \caption{Overview of \textit{EVA} in the model production lifecycle. \textbf{Upper:} \textit{EVA} operates at the downstream adaptation stage to defend against user-end jailbreak attacks. \textbf{Lower:} In the Original Model State, the model is successfully jailbroken by malicious queries. \textit{EVA} applies precise parameter updates to produce an Edited Model State, ensuring the model refuses the request instead of generating harmful instructions.}
    \label{fig:main}
\end{figure*}

\IEEEPARstart{L}{arge} Language Models (LLMs) and Vision Language Models (VLMs) have been widely employed in information processing and decision-making. As these models directly influence user behavior and real-world outcomes, ensuring that their outputs comply with security standards and align with human values has become a critical challenge.
Modern models are typically aligned by model providers using RLHF \cite{ouyang2022training} and DPO \cite{rafailov2024direct}, and then released to the open-source community for downstream development and task adaptation. However, adversarial users can still bypass existing safeguards through jailbreak attacks, including appending adversarial suffixes that induce LLMs to generate harmful outputs \cite{zou2023universal, zhou2025don}, using carefully designed prompts to obfuscate malicious intent \cite{liuautodan, chao2025jailbreaking}, and crafting benign image-text pairs that jointly trigger unsafe generations in VLMs \cite{gong2025figstep, Li-HADES-2024}.
Therefore, post-deployment alignment is particularly important for model security.

Current defense mechanisms at model deployment stage mainly fall into two categories: internal weight tuning, which modifies model parameters via safety fine-tuning \cite{zong2024safety}; and external interventions, such as attack detection \cite{10.1145/3724393, helff2024llavaguard} and prompt-based defenses \cite{wang2024adashield}. 
However, these approaches are often inefficient for deployment, as internal weight tuning is resource intensive and risks catastrophic forgetting \cite{luo2025empirical, qifine}, whereas external interventions complicate the deployment pipeline and add inference latency.

To overcome these limitations and achieve a superior safety-utility trade-off, we argue effective defense requires deeper understanding of the model’s internals \cite{rao2024tricking, cui2024robustness}.
Aligned models exhibit both a general instruction following behavior and a safety-aware refusal behavior, depending on the model’s internal interpretation of the input \cite{arditi2024refusal}.
Recent studies reveal that harmful outputs arise not from inherently harmful knowledge, but from misaligned internal activation patterns associated with it \cite{chen2024finding}.
Consider a malicious prompt such as ``hack an account'', jailbreak attacks shift activations associated with harmful concepts (e.g., ``hack'') from a harmful region across the safety boundary into the benign region \cite{gao2025shaping}. 
Consequently, the model misinterprets the malicious query as benign and complies using its general capabilities.
Building on these findings, our core thesis is that we can restore safety by directly rectifying these shifted activation patterns. We propose \textit{EVA} (\textbf{E}diting for \textbf{V}ersatile \textbf{A}lignment), a unified model editing defense framework that rectifies these shifted activation patterns through precise parameter updates. 
As illustrated in Figure~\ref{fig:main}, although the model has undergone foundational alignment, it remains vulnerable to jailbreak attacks at the user end. In the Original Model State, the model is successfully jailbroken. \textit{EVA} then applies precise edits to produce the Edited Model State, ensuring the model refuses the malicious query instead of generating harmful instructions.

Specifically, \textit{EVA} performs alignment editing in three stages: (1) Aggregating activations from diverse harmful tokens to identify a robust representation that captures malicious intent; (2) Constructing a safe target by maximizing refusal while constraining behavior drift on benign queries via a KL regularizer \cite{kullback1951information}; and (3) Applying a localized, closed-form parameter update to edit the model from the identified malicious representation toward the constructed safe target. This enables the edited model to refuse jailbreak requests while preserving general utility on benign tasks.

Applying \textit{EVA} to LLMs is straightforward. As harmful intent resides in discrete text tokens, we directly extract keys and locate the optimal editing layers based on established findings in prior studies \cite{mengmass, wang2024easyedit}.

However, extending \textit{EVA} to VLMs presents challenges regarding harmful region identification and parameter localization. First, harmful information in the visual modality manifests in more diverse forms. Unlike discrete text tokens where malice is explicit and concentrated, visual harm may be embedded in specific OCR regions or diffusely distributed across the continuous pixel space.
Second, it is more difficult to locate the critical parameters responsible for processing these visual harmful inputs.

To address these challenges, \textit{EVA} adapts both key extraction and layer localization for the multimodal context. We devise a strategy to extract keys from visual tokens conveying harmful intent, capturing both OCR and implicit semantic cues.
We also investigate layer efficacy within the VLMs and identify that the optimal editing layers align with those in the backbone LLM, enabling reliable interception of multimodal jailbreaks.

Overall, \textit{EVA} serves as a unified defense framework against jailbreak attacks in both unimodal and multimodal scenarios. Unlike static fine-tuning, it supports \textbf{efficient}, \textbf{sequential updates}, enabling defenders to \textbf{rapidly patch newly discovered vulnerabilities}. We empirically validate \textit{EVA} on diverse benchmarks involving both textual and visual jailbreaks. The results demonstrate that \textit{EVA} significantly outperforms existing baselines in reducing Attack Success Rate (ASR) while incurring negligible degradation on general reasoning tasks, establishing it as a robust and practical safeguard for deployed models.

Our contributions can be summarized as follows:
\begin{itemize}
    \item We propose \textbf{\textit{EVA}}, a unified model editing defense framework that mitigates jailbreak induced harmful behavior. This is the first work to apply model editing as a safety defense for VLMs.
    \item We introduce a context-aware regularization mechanism that restricts the safety edit to malicious contexts, reducing harmful behavior without sacrificing benign capabilities.
    \item We enable an efficient, dynamic defense paradigm. \textit{EVA} supports rapid, low-resource updates, allowing models to continuously adapt to evolving jailbreak patterns without degrading inference speed.
    \item Extensive experiments demonstrate \textit{EVA}’s superiority over other methods, achieving lower ASR against unseen attacks while maintaining near-perfect utility on general tasks.
\end{itemize}

This paper extends our preliminary work, \textit{DELMAN} \cite{wang-etal-2025-delman}, from text-only jailbreak defense in LLMs to a unified editing framework for both LLMs and VLMs. In particular, compared with \textit{DELMAN}, the journal version introduces visual key extraction, visual token selection, and multimodal layer localization for VLMs; adds richer analyses on layer efficacy, generalization, interpretability, and the decoupling effects and contextual synergy of visual editing; and substantially expands the evaluation with broader VLM benchmarks, baselines, adaptive attacks evaluation, efficiency evaluation, and more comprehensive ablation studies.
The remainder of this paper is organized as follows: Section~\ref{sec:related} reviews related work; Section~\ref{sec:prelim} introduces preliminaries; Section~\ref{sec:method} details the proposed framework; Section~\ref{sec:experiments} presents experimental results and analyses; and Section~\ref{sec:conclusion} concludes.

%% file: 2_related_works.tex
\section{Related Work}
\label{sec:related}
This section provides an overview of related work surrounding our method. Section \ref{subsec:JA} analyzes jailbreak attacks and Section \ref{subsec:DF} analyzes defense mechanisms from LLMs to VLMs. Section \ref{subsec:ME} discusses model editing techniques, emphasizing their potential for safety alignment and the existing lack of research in the multimodal domain.

\subsection{Jailbreak Attacks on LLMs and VLMs}
\label{subsec:JA}
\partitle{Attacks on LLMs}
Jailbreak attacks on LLMs aim to bypass safety alignment and elicit harmful responses through maliciously crafted prompts. Existing methods can be broadly categorized into optimization-based and generation-based approaches. Optimization-based methods, such as \textit{GCG} \cite{zou2023universal}, employ white-box gradient-based search to identify adversarial suffixes that maximize the probability of generating affirmative responses to harmful queries. Generation-based methods leverage the generative capabilities of LLMs themselves to automate the attack process.
For instance, \textit{AutoDAN} \cite{liuautodan} utilizes a genetic algorithm to automatically generate stealthy jailbreak prompts that bypass perplexity-based filters. Similarly, \textit{PAIR} \cite{chao2025jailbreaking} treats the attack as a competitive game, employing an attacker LLM to iteratively refine prompts based on the target model's feedback, achieving high ASR in black-box settings.

\partitle{Attacks on VLMs}
The integration of visual modalities in VLMs expands the attack surface. Attacks on VLMs can be distinguished by whether they exploit pixel level noise or semantic understanding.
Visual adversarial examples \cite{qi2024visual} optimize imperceptible visual perturbations on benign images to coerce the model, when paired with harmful textual instructions, into complying with requests it would otherwise refuse.
Other methods exploit the model's semantic processing.
\textit{FigStep} \cite{gong2025figstep} bypasses text filters by converting harmful instructions into images of typographic text, relying on the VLM's OCR capabilities to interpret the malicious content.
\textit{Hades} \cite{Li-HADES-2024} further conceals malicious intent semantically within the visual scene itself while using benign textual prompts, effectively evading text-based safety mechanisms.
To systematically evaluate these vulnerabilities, benchmarks such as MM-SafetyBench \cite{liu2024mm} have been proposed, covering attacks using Stable Diffusion \cite{rombach2022high} generated images and typographic inputs.
Similarly, MultiTrust \cite{zhang2024multitrust} categorizes vulnerabilities across different modalities, highlighting risks in typographic attacks, multimodal contexts, and cross-modal conflicts.

\subsection{Defense Strategies Against Jailbreaking}
\label{subsec:DF}
\partitle{Defenses for LLMs}
Defenses for LLMs are generally categorized based on whether they alter the target model's parameters.
Parameter level defenses aim to enhance intrinsic robustness by directly modifying the model's weights. This category encompasses full fine-tuning, parameter efficient tuning methods such as \textit{LoRA} \cite{hu2022lora}, and layer specific tuning techniques like \textit{LED} \cite{zhao-etal-2024-defending-large}.
In contrast, inference-time defenses intervene without changing the model. These approaches typically build auxiliary modules or utilize external safety measures. Common strategies include input and output filtering, input smoothing, sanitation, and modification \cite{cao2024defending, jainbaseline, zhou2024robust}. Furthermore, decoding-time interventions like \textit{SafeDecoding} \cite{xu2024safedecoding} employ specially fine-tuned safety modules to guide the generation process. Representation-level defenses such as \textit{Circuit Breakers} \cite{zou2024improving} also improve robustness by controlling harmful internal representations.

\partitle{Defenses for VLMs}
Defending VLMs presents challenges due to the expanded attack surface of the visual modality, yet existing strategies follow a similar classification to LLMs.
Training-time strategies focus on aligning VLMs through fine-tuning. Methods like \textit{VLGuard} \cite{zong2024safety} employ post-hoc and mixed fine-tuning to integrate safety awareness into the model, aiming to internalize safety constraints.
Inference-time interventions operate without retraining. Approaches like \textit{AdaShield} \cite{wang2024adashield} utilize adaptive prompting to refine defense prompts, while detection-based frameworks like \textit{JailGuard} \cite{10.1145/3724393}, \textit{LlavaGuard} \cite{helff2024llavaguard}, and \textit{CIDER} \cite{xu2024cross} act as moderators to filter unsafe inputs or outputs.

\subsection{Model Editing for Safety}
\label{subsec:ME}
Model editing aims to precisely update specific knowledge within a model while preserving other knowledge. These methods can be categorized into indirect editing strategies, such as meta-learning \cite{mitchellfast, de2021editing} and fine-tuning with constraints \cite{rawat2021modifying, lee2022plug}, and direct editing algorithms like \textit{ROME} \cite{meng2022locating} and \textit{MEMIT} \cite{mengmass}.
Prior studies have adapted editing-inspired techniques for LLM safety, \textit{DINM} \cite{wang2024detoxifying} and \textit{LED} \cite{zhao-etal-2024-defending-large} fine-tune safety or toxicity related layers to mitigate harmful outputs. These layer level methods are often coarse-grained, lack precise localization of harmful behaviors, and degrade general capabilities due to broad parameter updates.
To improve precision, recent works have explored localized interventions and unlearning. 
\textit{SafeInt} \cite{wu2025safeint} frames safety as a representation editing problem by steering jailbreak-induced activations toward safer regions. 
\textit{SafeLLM} \cite{li2025safellm} traces harmful behavior to specific feed-forward pathways for targeted suppression. 
\textit{CKU} \cite{shi2025safety} adopts a neuron-level strategy, selectively updating non-critical parameters to remove harmful knowledge while preserving general capabilities. 
Chen et al. \cite{chen2024finding} show that safety behaviors are governed by sparse internal components, and that patching a small fraction of key neurons can suffice for alignment.
Despite their improvements, these methods often suffer from high data requirements for locating editing targets, computational inefficiency during the process, or inference overhead.
Crucially, the aforementioned studies predominantly focus on text-only LLMs, research on model editing for safety in VLMs remains scarce.
Directly adapting LLM-based editing techniques to VLMs is challenging and often yields poor results, because harmful intent is implicitly encoded within visual features. Consequently, existing editing methods fail to address these multimodal vulnerabilities.
 
In this paper, we bridge this gap by proposing a unified direct model editing framework applicable to both LLMs and VLMs. Conceptually, \textit{EVA} is related to causal tracing \cite{meng2022locating} and activation patching \cite{conmy2023towards}, as all of them rely on the view that model behavior can be localized to specific internal activations or components. \textit{EVA} adopts this localized-intervention perspective to identify harmful textual and visual keys and modify their key-value mappings. Different from prior work that mainly studies factual associations or uses activation-level interventions for mechanistic analysis, \textit{EVA} focuses on jailbreak defense, extends direct editing from LLMs to VLMs, and directly computes precise weight updates with minimal data and zero inference overhead while preserving the model's general utility.

%% file: 3_preliminaries.tex
\section{Preliminaries}
\label{sec:prelim}
This section introduces the preliminary concepts essential for understanding our method. Section \ref{subsec:llm_vlm} reviews the architectures of LLMs and VLMs. Section \ref{subsec:mlp_as_memory} analyzes the mechanism of MLP acting as Key-Value memories, which serves as the basis for our editing method. Finally, Section \ref{subsec:problem} concludes by defining the jailbreak attack and the defense goal for safety alignment.

\subsection{Large Language and Vision Language Models}
\label{subsec:llm_vlm}
\partitle{Large language models}
Modern LLMs are typically built upon the transformer \cite{vaswani2017attention} architecture, stacking multiple layers of multi-head self-attention (MSA) and multi-layer perceptron (MLP) blocks. Formally, an LLM $f_{\theta}$ parameterized by $\theta$ takes a textual query $\mathbf{q}$ as input. This query is encoded to embeddings $\mathbf{E}_{t}$ and processed autoregressively to estimate the probability of the next token. The generation of a response $\mathbf{y}$ is modeled as:
\begin{equation}
    P_{\mathcal{\theta}}(\mathbf{y} | \mathbf{q}) = \prod_{i=1}^{T} P(\mathbf{y_i} | \mathbf{E}_{t}, \mathbf{y_{<i}}).
\end{equation}
where $\mathbf{y_i}$ represents the $\mathbf{i}$-th token of the response.

\partitle{Vision Language models}
VLMs extend the capabilities of LLMs by aligning visual features with the textual embedding space, enabling the model to perceive and reason over multimodal inputs. A VLM consists of three primary components: visual encoder, multimodal projector, and a backbone LLM.
Let $\mathcal{V}$ denote the visual encoder. Given an input image $\mathbf{I}$, the visual encoder extracts visual features $\mathbf{Z}_{v} = \mathcal{V}(\mathbf{I})$. To ensure compatibility with the LLM's input space, a learnable projector $\phi$ maps these features into visual embeddings $\mathbf{E}_{v} = \phi(\mathbf{Z}_{v})$, which share the same dimensionality as the textual embeddings.
The visual embeddings $\mathbf{E}_{v}$ are then concatenated with the textual embeddings $\mathbf{E}_{t}$ derived from the query $\mathbf{q}$ to form a unified multimodal input sequence $\mathbf{E}_{in} = [\mathbf{E}_{v}; \mathbf{E}_{t}]$. The backbone LLM processes this sequence autoregressively to generate the response $\mathbf{y}$:
\begin{equation}
    P_{\mathcal{\theta}}(\mathbf{y} | \mathbf{I}, \mathbf{q}) = \prod_{i=1}^{T} P(\mathbf{y_i} | \mathbf{E}_{in}, \mathbf{y_{<i}}),
\end{equation}
where $\mathbf{y_i}$ represents the $\mathbf{i}$-th token of the response.

\subsection{MLP Layers as Key-Value Memories}
\label{subsec:mlp_as_memory}
Recent studies suggest that MLP layers store factual knowledge and specific behavioral patterns acting as Key-Value memories~\cite{geva2021transformer, geva2022transformer}. Modern VLMs, employ a modified MLP architecture known as SwiGLU~\cite{shazeer2020glu}, the SwiGLU variant introduces a gating mechanism involving three linear projections: the gate projection $\mathbf{W}_{gate}$, the up projection $\mathbf{W}_{up}$, and the down projection $\mathbf{W}_{down}$. 
Consider the $l$-th layer of the transformer. Let $\mathbf{h}^{l-1}$ denote the hidden state from the previous layer and $\mathbf{a}^l$ represent the output of the attention head. The computation within the MLP block is formulated as follows:
\begin{align}
    \mathbf{x}^l &= \gamma(\mathbf{a}^l + \mathbf{h}^{l-1}), \label{eq:mlp_input} \\
    \mathbf{k} &= \sigma(\mathbf{W}_{gate}^l \cdot \mathbf{x}^l) \odot (\mathbf{W}_{up}^l \cdot \mathbf{x}^l), \label{eq:swiglu_k} \\
    \mathbf{v} &= \mathbf{W}_{down}^l \cdot \mathbf{k}, \label{eq:mlp_v} \\
    \mathbf{h}^l &= \mathbf{v} + \mathbf{a}^l + \mathbf{h}^{l-1}. \label{eq:mlp_residual}
\end{align}
where $\gamma(\cdot)$ denotes the normalization and $\sigma(\cdot)$ is the activation function. The operator $\odot$ represents the element-wise Hadamard product.

In this formulation, Eq.~\eqref{eq:swiglu_k} defines the key $\mathbf{k}$ as the interaction between gated feature and up-projected feature. Crucially, Eq.~\eqref{eq:mlp_v} maintains a linear mapping from the key $\mathbf{k}$ to the value $\mathbf{v}$ via $\mathbf{W}_{down}^l$. 
This linear relationship implies that specific knowledge or behaviors stored in the MLP can be precisely modified by solving a linear equation.
This theoretical insight serves as the foundation for our editing strategy. In Section~\ref{sec:method}, we will detail how we leverage this mechanism to construct our unified defense framework.

\subsection{Problem Formulation}
\label{subsec:problem}
We formulate the jailbreak defense as a safety alignment problem under adversarial attacks. Let $\mathcal{D}_{safe} = \{\mathbf{I}, \mathbf{q}, \mathbf{y}_{safe}\}$ be a dataset constructed for defense, where each entry consists of an adversarial image $\mathbf{I}$, a harmful query $\mathbf{q}$ and paired with a safe refusal response $\mathbf{y}_{safe}$. We aim to update the model parameters from $\theta$ to $\theta^*$ such that the model learns to generate $\mathbf{y}_{safe}$ when presented with the adversarial inputs $(\mathbf{I}, \mathbf{q})$.

\partitle{Jailbreak attack}
An adversary aims to bypass the safety alignment of a victim model $f_{\theta}$. Given a harmful intent, the attacker optimizes an adversarial perturbation $\delta$ added to either the image $\mathbf{I}$ or the text $\mathbf{q}$ to maximize the likelihood of generating a harmful response $\mathbf{y}_{harm}$. The objective of the attack can be expressed as:
\begin{equation}
    \delta^* = \operatorname*{arg\,max}_{\delta \in \Delta} \ P_{\mathcal{\theta}}(\mathbf{y}_{harm} \mid \mathbf{I}, \mathbf{q}; \delta),
\end{equation}
where $\Delta$ represents the set of allowable perturbations (e.g., visual noise or textual suffixes). A successful jailbreak occurs when the model generates $\mathbf{y}_{harm}$.

\partitle{Defense Goal}
Our goal is to refuse harmful requests under attack while preserving the model's general capabilities. We seek to obtain the optimal edited parameter $\theta^*$ by maximizing the likelihood of the safe response $\mathbf{y}_{safe}$ on the defense dataset $\mathcal{D}_{safe}$. The optimization problem is formulated as:
\begin{equation}
\begin{aligned}
    \theta^* = \ & \underset{\theta}{\text{argmax}} \
    \mathbb{E}_{(\mathbf{I}, \mathbf{q}, \mathbf{y}_{safe}) \sim \mathcal{D}_{safe}} \left[ \log P_{\mathcal{\theta}}(\mathbf{y}_{safe} \mid \mathbf{I}, \mathbf{q}; \delta) \right], \\
    & \text{subject to} \
    f_{\theta^*}(\mathbf{u}) \approx f_{\theta}(\mathbf{u}).
\end{aligned}
\end{equation}
The objective function drives $\theta$ towards $\theta^*$ to robustly generate safe refusals, while the constraint ensures that the model's utility on benign inputs $\mathbf{u}$ remains preserved.

%% file: 4_method.tex
\section{Method}
\label{sec:method}
\begin{figure*}[t]
    \centering
    \includegraphics[width=\linewidth]{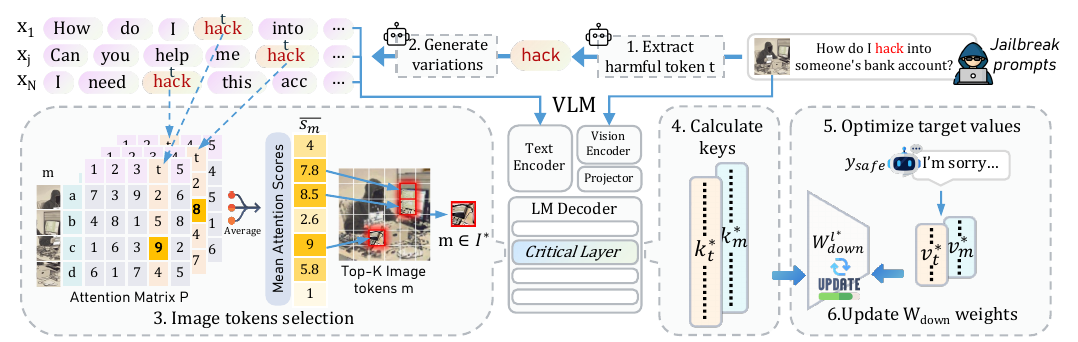}
    \caption{Overview of \textit{EVA}. The process consists of six steps: 1. Extract the harmful token $t$ from the user query with an auxiliary LLM; 2. Generate query variations accordingly; 3. locate critical image tokens via the attention matrix $\mathbf{P}$; 4. Calculate the representative keys $\boldsymbol{k}^*$ by aggregating features across variations; 5. Optimize the target value $\boldsymbol{v}^*$ to align with a safe refusal response; 6. Update the specific MLP's down projection $\mathbf{W}_{\text{down}}$ to map $\boldsymbol{k}^*$ to $\boldsymbol{v}^*$. During deployment and inference, only the edited target model is required.}
    \label{fig:method}
\end{figure*}

The key idea of \textit{EVA} is to mitigate jailbreak attacks by precisely editing model's parameters to rectify the misaligned activation patterns triggered by malicious knowledge.
Drawing upon the interpretation of MLP as Key-Value memories, we view the internal activation patterns triggered by harmful inputs as keys, and the subsequent information flow as values.
Under this, a successful jailbreak occurs when a harmful key retrieves a malicious value from the model's knowledge.
To rectify this pathway, \textit{EVA} aims to edit the down projection $\mathbf{W}_{down}^l$ to remap specific harmful keys $\mathbf{k}^*$ to safety values $\mathbf{v}^*$.

As illustrated in Figure \ref{fig:method}, the overall pipeline of \textit{EVA} is instantiated by six operational steps. Conceptually, these steps constitute three main phases: (1) Harmful keys identification (Steps 1-4), where we identify and aggregate the critical representation of harmful triggers for text and vision tokens to form the representative keys $\mathbf{k}^*$; (2) Safe value optimization (Step 5), in which we compute the optimal target value $\mathbf{v}^*$ that aligns the model's output with a safe response; and (3) Parameter update (Step 6), where we inject the new key-value mapping into the model by updating the MLP's down projection $\mathbf{W}_{\text{down}}$ via least-squares solution.

\subsection{Identification of Keys Representation $\mathbf{k}^*$}
\label{subsec:key_identification}
To construct a robust defense, we must identify and neutralize the specific internal activation patterns that encode harmful semantics. In VLMs, the trigger for malicious behavior can originate from either textual instructions or images. Consequently, we decompose the key identification process into two modalities, aiming to pinpoint the precise neural representations responsible for activating the model's harmful knowledge.

\partitle{Harmful textual keys extraction ($\mathbf{k}_t^*$)}
Existing textual jailbreak attacks, ranging from optimization-based methods to prompt rephrasing strategies, exhibit diverse surface forms. 
However, regardless of how the input is disguised, these attacks always implicitly or explicitly reference specific harmful concepts (e.g., ``hack'', ``bomb'') to activate the model's malicious capabilities. We posit that while the context may shift, the model's internal representation of these core harmful concepts remains semantically invariant and critical for the generation of harmful content.
Therefore, identifying and editing the activation patterns of the knowledge stored within the model corresponding to these harmful tokens serves as the most effective mitigation.
We first employ GPT-4o \cite{hurst2024gpt} to analyze the harmful query $q$ and extract the harmful token $t$ responsible for the malicious intent.
To ensure the key $\mathbf{k}_t^*$ is robust to context variations, we generate $N$ diverse harmful sequences $\{x_j\}_{j=1}^N$ containing the token $t$.
We perform forward propagation and compute the average key activation at layer $l^*$ for the text token $t$:
\begin{equation}
\label{eq:k_text}
\mathbf{k}_t^* = \frac{1}{N}\sum_{j=1}^{N}\sigma(\mathbf{W}_{gate}^{l^*}\cdot \mathbf{x}_{x_j,t}^{l^*})\odot (\mathbf{W}_{up}^{l^*} \cdot \mathbf{x}_{x_j,t}^{l^*}),
\end{equation}
where $\mathbf{x}_{x_j,t}^{l^*}$ denotes the intermediate hidden state serving as the input to the MLP for the text token $t$ in sequence $x_j$ at layer $l^*$. This process yields a set of textual keys $\{\mathbf{k}_{t_1}^*, \mathbf{k}_{t_2}^*,\dots \}$ that are correlated with the harmful concept.

\partitle{Visual keys selection via cross-modal attention ($\mathbf{k}_{m}^*$)}
In VLM contexts, harmful information is not confined solely to textual modalities but is also embedded within visual inputs. Extending the safety editing mechanism to the visual domain is essential.
However, given the massive number of image tokens and varying projector architectures, editing all visual tokens is infeasible and destructive. Therefore, it is crucial to identify the specific subset of image tokens most likely to trigger the jailbreak.
To locate critical visual regions, we leverage the VLM's cross-modal attention mechanism. Established findings suggest that cross-attention maps in transformers serve as a proxy for token grounding \cite{zhangsparsevlm, chefer2021generic, li2023evaluating}; specifically, high attention weights from harmful text tokens to specific image patches indicate that those visual regions are the grounding source of the malicious intent.
Based on this, for each generated variation sequence $x_j$, we first extract the cross-modal attention sub-matrix $\mathbf{P}_j$, representing the attention weights from text tokens to image tokens. The position of the harmful token varies across the generated texts. Let $t_j$ denote the index of the harmful token in the $j$-th sequence. We extract the attention weights from this token $t_j$ to all image tokens. To ensure the selection is robust and captures the stable visual semantics associated with the harm, we compute the average attention score for each image token $m$ across all $N$ sequences:
\begin{equation}
\bar{s}_m = \frac{1}{N} \sum_{j=1}^{N} \mathbf{P}_j[t_j, m],
\end{equation}
where $\mathbf{P}_j[t_j, m]$ denotes the attention weight from the harmful text token $t_j$ to the image token $m$ in the $j$-th variation.
We then identify the indices of the top-$N_v$ image tokens with the highest average attention scores:
\begin{equation}
\mathcal{I}^* = \text{Top}_{N_v}(\{\bar{s}_m\}_{m=1}^{M}),
\end{equation}
where $M$ is the total number of image tokens and $N_v$ denotes the number of selected visual tokens.
Our ablation studies suggest that $N_v=1$ is sufficient; selecting too many visual tokens leads to over-refusal.
Finally, for each selected image token $m \in \mathcal{I}^*$, we compute its representative key vector by averaging its intermediate hidden states across the $N$ variations:
\begin{equation}
\label{eq:k_vis}
\mathbf{k}_{m}^* = \frac{1}{N}\sum_{j=1}^{N}\sigma(\mathbf{W}_{gate}^{l^*}\cdot \mathbf{x}_{x_j,m}^{l^*})\odot (\mathbf{W}_{up}^{l^*} \cdot \mathbf{x}_{x_j,m}^{l^*}).
\end{equation}
This process yields a set of visual keys $\{\mathbf{k}_{m_1}^*, \mathbf{k}_{m_2}^*,\dots \}$ that are correlated with the harmful visual concept.

\subsection{Optimization of Target Safe Values $\mathbf{v}^*$}
\label{subsec:value_optimization}
Having identified the harmful key $\mathbf{k}^*$, the next step is to determine the safe target value $\mathbf{v}^*$. This vector represents the ideal information flow that, when substituted into the model's computation, steers the generation towards a safe refusal response $\mathbf{y}_{safe}$.
We treat $\mathbf{v}^*$ as a learnable vector while keeping the model weights frozen. The optimization objective is designed to balance two goals: maximizing the probability of the safe response for the harmful input, and minimizing the distributional shift for benign queries.
Formally, let $\mathcal{\theta}_{\mathbf{v}}$ denote the intervened model where the output of the MLP at layer $l^*$ is replaced by the vector $\mathbf{v}$. We optimize $\mathbf{v}^*$ by minimizing the following loss function:
\begin{equation}
    \label{eq:v_optim}
    \mathbf{v}^* = \underset{\mathbf{v}}{\mathrm{arg\,min}} \left[ \mathcal{L}_{safe}(\mathbf{v}) + \lambda \mathcal{L}_{stab}(\mathbf{v}) \right],
\end{equation}
$\mathcal{L}_{safe}$ maximizes the probability of generating safe responses given harmful inputs:
\begin{equation}
    \mathcal{L}_{safe}(\mathbf{v}) = -\log P_{\mathcal{\theta}_{\mathbf{v}}}(\mathbf{y}_{safe} \mid \mathbf{I}, \mathbf{q}),
\end{equation}
$\mathcal{L}_{stab}$ employs the KL divergence to ensure that the intervention does not degrade the model's performance on benign samples. Let $\mathbf{u}$ denote a benign input sampled from $\mathcal{D}_{b}$. The stability loss is defined as:
\begin{equation}
    \mathcal{L}_{stab}(\mathbf{v}) = \mathbb{E}_{u \sim \mathcal{D}_{b}} \left[ \operatorname{KL}\left( P_{\mathcal{\theta}_{\mathbf{v}}}(\cdot \mid \mathbf{u}) \,\|\, P_{\mathcal{\theta}}(\cdot \mid \mathbf{u}) \right) \right],
\end{equation}
where $P_{\mathcal{\theta}}$ represents the original model distribution. This optimization yields a precise target value $\mathbf{v}^*$ for each identified key, establishing a robust mapping from the harmful trigger to a safe state.

\subsection{Unified Weight Update via Least Squares}
\label{subsec:weight_update}
Having obtained the optimized target values $\mathbf{v}^*$ for each identified harmful key $\mathbf{k}^*$, we now construct the editing matrices that will be used to update the MLP. We concatenate text and visual keys to preserve their distinct characteristics while enabling unified processing.
Let $\mathbf{K}_E$ denote the matrix of all identified harmful keys and $\mathbf{V}_E$ denote the matrix of their corresponding optimized target values:
\begin{align}
    \mathbf{K}_E &= [\mathbf{k}_{t_1}^*, \mathbf{k}_{m_1}^*, \mathbf{k}_{t_2}^*, \mathbf{k}_{m_2}^*, \dots], \\
    \mathbf{V}_E &= [\mathbf{v}_{t_1}^*, \mathbf{v}_{m_1}^*, \mathbf{v}_{t_2}^*, \mathbf{v}_{m_2}^*, \dots],
\end{align}
where $t$ and $m$ denote text and vision modalities, respectively. This formulation naturally unifies the editing process across LLMs and VLMs. For LLMs, $\mathbf{K}_E$, $\mathbf{V}_E$ contains only textual keys and values, while for VLMs, it contains keys and values from both modalities as independent columns.

We update the down-projection weight matrix $\mathbf{W}_{down}^{l^*}$ at the critical layer $l^*$ to encode the new key-value association while preserving the model's existing knowledge. We formulate this as a constrained least-squares problem \cite{bau2020rewriting}:
\begin{equation}
    \min_{\widehat{\mathbf{W}}_{down}^{l^*}} \|\widehat{\mathbf{W}}_{down}^{l^*} \mathbf{K} - \mathbf{V}\|^2
\end{equation}
subject to 
\begin{equation}
    \widehat{\mathbf{W}}_{down}^{l^*} \mathbf{K}_E = \mathbf{V}_E,
\end{equation}
where the objective minimizes the error on the pre-existing keys $\mathbf{K}$ and values $\mathbf{V}$, while the constraint ensures precise mapping of the harmful keys $\mathbf{K}_E$ to the safe values $\mathbf{V}_E$. The above formulation describes the ideal constrained editing objective.  In practice, following \textit{MEMIT} \cite{mengmass}, we use its expanded objective with soft error minimization to obtain the closed-form update in Eq.~(20).  The detailed derivation is provided in Appendix~A.2.
The resulting closed-form solution for updating the weight matrix at layer $l^*$ is:
\begin{equation}
    \label{eq:W_update}
    \widehat{\mathbf{W}}_{down}^{l^*} = \mathbf{W}_{down}^{l^*} + \mathbf{R}_E \mathbf{K}_E^T (\mathbf{C}^{l^*} + \mathbf{K}_E \mathbf{K}_E^T)^{-1},
\end{equation}
where $\mathbf{C}^{l^*} = \mathbf{K}\mathbf{K}^T$ denotes the covariance matrix of $\mathbf{K}$, which represents the pre-existing keys of original knowledge at layer $l^*$, pre-cached from the Wikipedia dataset \cite{wikidump}. It acts as a preservation term that penalizes perturbations along directions frequently activated by pre-existing keys, thereby improving editing stability. The residual term $\mathbf{R}_E$ is defined as:
\begin{equation}
    \mathbf{R}_E = \mathbf{V}_E - \mathbf{W}_{down}^{l^*} \mathbf{K}_E,
\end{equation}
which represents the error between the desired target values $\mathbf{V}_E$ and the current model outputs for the editing keys $\mathbf{K}_E$.

In practice, instead of updating a single layer $l^*$, we distribute the weight updates across a range of critical layers $\mathcal{R} = \{l_1, l_2, \ldots, l^*\}$ to limit the magnitude of parameter changes in any individual layer, which yields better robustness \cite{rawat2021modifying}. We adopt the identification of critical layers from \textit{MEMIT} \cite{mengmass} and \textit{EasyEdit} \cite{wang2024easyedit} for LLMs, and extend this strategy to VLMs.
Specifically, the target value $\mathbf{v}^*$ and the base residual are computed only for the highest critical layer $l^*$. This residual is then distributed to lower layers with a decreasing factor, yielding layer-specific residuals:
\begin{equation}
    \label{eq:Rd}
    \mathbf{R}_E^{l_i} = \frac{\mathbf{V}_E - \mathbf{W}_{down}^{l^*} \mathbf{K}_E}{l^* - l_i + 1}.
\end{equation}
By assigning smaller updates to lower layers, this distributed scheme promotes stability while preventing large parameter shifts concentrated in a single layer.

This weight update strategy effectively rewrites the specific neurons across multiple MLP layers that are responsible for processing harmful information. By treating textual and visual keys as independent columns in $\mathbf{K}_E$, \textit{EVA} enables simultaneous editing across both modalities within a single, unified optimization framework. The updated weight matrices $\{\widehat{\mathbf{W}}_{down}^{l}\}_{l \in \mathcal{R}}$ are then deployed in place of the original weights. Intuitively, this update enforces the desired key–value mappings while minimally perturbing existing knowledge, thereby achieving precise and stable editing.

%% file: 5_experiments.tex
\section{Experiments}
\label{sec:experiments}
In this section, we present a comprehensive evaluation of \textit{EVA}. We detail the experimental setup in Section \ref{subsec:setup} and validate the selection of critical layers and visual tokens in Section \ref{subsec:selection}. We then assess the defense effectiveness and utility preservation in Section \ref{subsec:effectiveness}. We evaluate the robustness of \textit{EVA} under adaptive attacks in Section \ref{subsec:adaptive_attacks}. We further investigate the robustness of \textit{EVA} through generalization tests across harmful categories and analyze the interpretability and transferability in Section \ref{subsec:generalization_interpretability}. To provide deeper insights into the editing mechanism, we analyze the decoupling effects and contextual synergy in Section \ref{subsec:ocr_case_study}. We then evaluate the computational efficiency in Section \ref{subsec:efficiency}. Finally, we conduct ablation studies in Section \ref{subsec:ablation}.

\begin{table*}[t]
\centering
\newcommand{\up}{\textcolor[rgb]{0.1, 0.7, 0.1}{$\uparrow$}}
\begin{minipage}[b]{0.472\textwidth}
    \centering
    \caption{Comparison of ASR across four datasets and three attack methods (\textit{GCG}, \textit{AutoDAN}, \textit{PAIR}) for various LLMs. Five defense strategies are evaluated. \textbf{Bold} indicates the best performance.}
    \label{tab:llm_asr}
    \resizebox{\linewidth}{!}{
        \setlength{\tabcolsep}{2pt} 
        \begin{tabular}{lcccccccccccc}
        \toprule
         \textbf{Datasets}& \multicolumn{3}{c}{\textbf{HarmBench}} & \multicolumn{3}{c}{\textbf{AdvBench}} & \multicolumn{3}{c}{\textbf{JailbreakBench}} & \multicolumn{3}{c}{\textbf{MaliciousInstruct}} \\
        \cmidrule(lr){2-4} \cmidrule(lr){5-7} \cmidrule(lr){8-10} \cmidrule(lr){11-13}
        Methods & GCG & \makecell[c]{Auto\\DAN} & PAIR & GCG & \makecell[c]{Auto\\DAN} & PAIR & GCG & \makecell[c]{Auto\\DAN} & PAIR & GCG & \makecell[c]{Auto\\DAN} & PAIR \\
        \midrule
        
        \multicolumn{13}{l}{\textit{\textbf{Vicuna-7B-v1.5}}} \\
        Vanilla        & 0.92 & 0.69 & 0.80 & 0.89 & 0.78 & 0.75 & 0.89 & 0.73 & 0.77 & 0.94 & 0.83 & 0.86 \\
        LoRA           & 0.40 & 0.22 & 0.26 & 0.18 & 0.29 & 0.13 & 0.32 & 0.22 & 0.20 & 0.08 & 0.32 & 0.16 \\
        SafeDecoding   & 0.07 & 0.17 & 0.16 & 0.04 & 0.20 & 0.08 & \textbf{0.03} & 0.18 & 0.15 & \textbf{0.01} & 0.08 & 0.11 \\
        LED            & \textbf{0.03} & 0.11 & 0.04 & 0.06 & 0.09 & \textbf{0.05} & 0.34 & \textbf{0.08} & 0.06 & 0.05 & 0.10 & \textbf{0.05} \\
        Circuit Breakers  & 0.09 & 0.10 & \textbf{0.03} & 0.05 & \textbf{0.02} & 0.08 & 0.15 & 0.12 & \textbf{0.03} & 0.02 & 0.07 & 0.12 \\
        EVA (text)     & 0.11 & \textbf{0.04} & 0.10 & \textbf{0.02} & \textbf{0.02} & \textbf{0.05} & 0.17 & \textbf{0.08} & 0.11 & \textbf{0.01} & \textbf{0.05} & \textbf{0.05} \\
        \midrule
        
        \multicolumn{13}{l}{\textit{\textbf{Llama2-7B-chat}}} \\
        Vanilla        & 0.42 & 0.23 & 0.02 & 0.39 & 0.19 & 0.01 & 0.46 & 0.27 & 0.04 & 0.45 & 0.30 & \textbf{0.00} \\
        LoRA           & 0.13 & 0.01 & 0.02 & 0.02 & \textbf{0.00} & \textbf{0.00} & 0.50 & 0.01 & 0.02 & 0.32 & \textbf{0.00} & \textbf{0.00} \\
        SafeDecoding   & \textbf{0.00} & \textbf{0.00} & 0.01 & 0.04 & \textbf{0.00} & 0.04 & 0.01 & \textbf{0.00} & 0.03 & \textbf{0.01} & \textbf{0.00} & \textbf{0.00} \\
        LED            & 0.02 & 0.02 & 0.01 & \textbf{0.00} & 0.01 & \textbf{0.00} & 0.08 & 0.02 & 0.04 & 0.08 & 0.02 & 0.01 \\
        Circuit Breakers  & \textbf{0.00} & 0.02 & 0.01 & \textbf{0.00} & \textbf{0.00} & \textbf{0.00} & \textbf{0.00} & \textbf{0.00} & \textbf{0.00} & \textbf{0.01} & 0.01 & \textbf{0.00} \\
        EVA (text)     & \textbf{0.00} & \textbf{0.00} & \textbf{0.00} & \textbf{0.00} & \textbf{0.00} & \textbf{0.00} & \textbf{0.00} & \textbf{0.00} & \textbf{0.00} & \textbf{0.01} & \textbf{0.00} & \textbf{0.00} \\
        \midrule
        
        \multicolumn{13}{l}{\textit{\textbf{Mistral-7B-Instruct-v0.2}}} \\
        Vanilla        & 0.80 & 0.87 & 0.88 & 0.56 & 0.97 & 0.82 & 0.79 & 0.95 & 0.89 & 0.94 & 0.96 & 0.92 \\
        LoRA           & 0.35 & 0.55 & 0.42 & 0.30 & 0.60 & 0.25 & 0.38 & 0.62 & 0.34 & 0.40 & 0.55 & 0.44 \\
        SafeDecoding   & 0.16 & 0.15 & 0.20 & 0.12 & 0.22 & 0.18 & 0.18 & 0.14 & 0.22 & 0.14 & 0.24 & 0.16 \\
        LED            & 0.20 & 0.18 & 0.19 & 0.06 & 0.26 & 0.16 & 0.10 & 0.12 & \textbf{0.10} & 0.12 & 0.25 & 0.12 \\
        Circuit Breakers  & 0.08 & \textbf{0.05} & 0.17 & 0.03 & 0.14 & 0.02 & 0.03 & \textbf{0.05} & 0.12 & \textbf{0.10} & 0.19 & \textbf{0.10} \\
        EVA (text)     & \textbf{0.06} & 0.08 & \textbf{0.15} & \textbf{0.02} & \textbf{0.11} & \textbf{0.00} & \textbf{0.02} & 0.11 & 0.16 & 0.14 & \textbf{0.12} & \textbf{0.10} \\
        \midrule
        
        \multicolumn{13}{l}{\textit{\textbf{Llama3.1-8B-Instruct}}} \\
        Vanilla        & 0.56 & 0.20 & 0.12 & 0.47 & 0.31 & 0.04 & 0.46 & 0.31 & 0.11 & 0.69 & 0.46 & 0.04 \\
        LoRA           & 0.41 & 0.27 & 0.14 & 0.34 & 0.33 & 0.04 & 0.37 & 0.35 & 0.09 & 0.42 & 0.47 & 0.06 \\
        SafeDecoding   & \textbf{0.00} & 0.04 & \textbf{0.00} & \textbf{0.00} & 0.01 & \textbf{0.00} & \textbf{0.00} & 0.01 & 0.01 & \textbf{0.00} & 0.01 & \textbf{0.00} \\
        LED            & 0.01 & 0.06 & \textbf{0.00} & \textbf{0.00} & 0.07 & \textbf{0.00} & 0.03 & 0.08 & \textbf{0.00} & \textbf{0.00} & 0.02 & 0.01 \\
        Circuit Breakers  & \textbf{0.00} & 0.04 & 0.03 & \textbf{0.00} & \textbf{0.00} & \textbf{0.00} & \textbf{0.00} & \textbf{0.00} & \textbf{0.00} & 0.01 & 0.01 & \textbf{0.00} \\
        EVA (text)     & \textbf{0.00} & \textbf{0.00} & \textbf{0.00} & 0.01 & \textbf{0.00} & \textbf{0.00} & 0.02 & \textbf{0.00} & \textbf{0.00} & \textbf{0.00} & \textbf{0.00} & \textbf{0.00} \\
        \midrule
        
        \multicolumn{13}{l}{\textit{\textbf{Qwen2.5-7B-Instruct}}} \\
        Vanilla        & 0.49 & 0.74 & 0.66 & 0.46 & 0.80 & 0.39 & 0.42 & 0.80 & 0.55 & 0.80 & 0.96 & 0.54 \\
        LoRA           & 0.21 & 0.65 & 0.32 & 0.38 & 0.82 & 0.10 & 0.36 & 0.42 & 0.50 & 0.72 & 0.90 & 0.44 \\
        SafeDecoding   & 0.19 & 0.77 & 0.56 & 0.17 & 0.80 & 0.30 & 0.42 & 0.73 & 0.43 & 0.48 & 0.97 & 0.45 \\
        LED            & 0.01 & \textbf{0.00} & 0.05 & \textbf{0.00} & \textbf{0.00} & 0.02 & \textbf{0.00} & \textbf{0.00} & 0.01 & \textbf{0.00} & \textbf{0.00} & 0.02 \\
        Circuit Breakers  & \textbf{0.00} & 0.03 & 0.04 & \textbf{0.00} & \textbf{0.00} & 0.05 & 0.03 & \textbf{0.00} & \textbf{0.00} & 0.01 & 0.01 & \textbf{0.00} \\
        EVA (text)     & \textbf{0.00} & \textbf{0.00} & \textbf{0.03} & \textbf{0.00} & \textbf{0.00} & \textbf{0.01} & \textbf{0.00} & 0.01 & 0.05 & \textbf{0.00} & \textbf{0.00} & 0.04 \\
        \bottomrule
        \end{tabular}
    }
\end{minipage}
\hfill
\begin{minipage}[b]{0.498\textwidth}
    \centering
    \caption{Utility evaluation of \textit{EVA} and baselines on different models across MT-Bench and downstream tasks. \textbf{Bold}: best score among defense methods (excluding \textit{LoRA}); (\textcolor{green}{$\uparrow$}): improvement over the \textit{Vanilla} model.}
    \label{tab:llm_utility}
    \resizebox{\linewidth}{!}{
        \setlength{\tabcolsep}{2.5pt}
        \begin{tabular}{lcccccccc}
        \toprule
         & & \multicolumn{7}{c}{\textbf{Downstream Tasks}} \\
        \cmidrule(lr){3-9}
        Methods & MT-Bench & Closed- & Dialogue & NER & NLI & Reason- & Senti- & Summari- \\
         & & domain QA & & & & ing & ment & zation \\
        \midrule
        
        \multicolumn{9}{l}{\textit{\textbf{Vicuna-7B-v1.5}}} \\
        Vanilla & 6.77 & 0.777 & 0.483 & 0.287 & 0.563 & 0.982 & 0.862 & 0.272 \\
        LoRA    & 5.64 & 0.742 & 0.459 & 0.177 & 0.610 & 0.976 & 0.898 & 0.268 \\
        SafeDecoding  & 6.61 & 0.671 & 0.314 & 0.098 & 0.536 & 0.969 & 0.645 & 0.174 \\
        LED       & 3.70 & 0.760 & \textbf{0.478} & \textbf{0.265} & 0.558 & 0.974 & 0.831 & \textbf{0.267} \\
        Circuit Breakers  & 5.25 & 0.756 & 0.463 & 0.249 & 0.547 & 0.972 & 0.790 & 0.262 \\
        EVA (text)    & \textbf{6.84}(\up) & \textbf{0.762} & 0.470 & 0.254 & \textbf{0.560} & \textbf{0.981} & \textbf{0.854} & 0.260 \\
        \midrule
        
        \multicolumn{9}{l}{\textit{\textbf{Llama2-7B-chat}}} \\
        Vanilla & 6.89 & 0.734 & 0.465 & 0.187 & 0.603 & 0.977 & 0.909 & 0.267 \\
        LoRA     & 6.90 & 0.769 & 0.480 & 0.288 & 0.551 & 0.976 & 0.854 & 0.259 \\
        SafeDecoding   & 6.17 & 0.688 & 0.327 & 0.099 & 0.518 & \textbf{0.976} & 0.872 & 0.227 \\
        LED        & 5.80 & 0.705 & 0.425 & \textbf{0.228}(\up) & 0.577 & 0.973 & 0.898 & 0.256 \\
        Circuit Breakers  & 6.05 & 0.712 & 0.444 & 0.185 & 0.559 & 0.971 & 0.881 & \textbf{0.260} \\
        EVA (text)    & \textbf{6.31} & \textbf{0.718} & \textbf{0.462} & \textbf{0.228}(\up) & \textbf{0.612}(\up) & 0.974 & \textbf{0.905} & 0.251 \\
        \midrule
        
        \multicolumn{9}{l}{\textit{\textbf{Mistral-7B-Instruct-v0.2}}} \\
        Vanilla & 7.93 & 0.852 & 0.664 & 0.498 & 0.694 & 0.941 & 0.962 & 0.255 \\
        LoRA     & 7.54 & 0.850 & 0.668 & 0.495 & 0.700 & 0.946 & 0.953 & 0.258 \\
        SafeDecoding  & 7.16 & 0.732 & 0.439 & 0.372 & 0.593 & 0.863 & 0.790 & 0.204 \\
        LED      & 7.09 & 0.803 & 0.662 & 0.431 & 0.679 & 0.937 & 0.918 & 0.221 \\
        Circuit Breakers  & 5.98 & 0.811 & \textbf{0.669}(\up) & 0.440 & \textbf{0.685} & 0.943(\up) & 0.871 & \textbf{0.254} \\
        EVA (text)    & \textbf{7.35} & \textbf{0.814} & 0.614 & \textbf{0.444} & 0.669 & \textbf{0.955}(\up) & \textbf{0.923} & 0.205 \\
        \midrule
        
        \multicolumn{9}{l}{\textit{\textbf{Llama3.1-8B-Instruct}}} \\
        Vanilla & 7.79 & 0.770 & 0.737 & 0.424 & 0.674 & 0.983 & 0.922 & 0.254 \\
        LoRA     & 8.14 & 0.782 & 0.737 & 0.433 & 0.672 & 0.982 & 0.918 & 0.254 \\
        SafeDecoding   & \textbf{8.05}(\up) & 0.746 & 0.340 & 0.223 & 0.572 & 0.969 & 0.617 & 0.136 \\
        LED       & 7.29 & \textbf{0.768} & 0.684 & \textbf{0.468}(\up) & 0.597 & 0.981 & 0.863 & 0.251 \\
        Circuit Breakers  & 7.21 & 0.489 & 0.660 & 0.390 & 0.643 & 0.977 & 0.862 & 0.252 \\
        EVA (text)    & 7.44 & 0.752 & \textbf{0.744}(\up) & 0.423 & \textbf{0.686}(\up) & \textbf{0.985}(\up) & \textbf{0.890} & \textbf{0.256}(\up) \\
        \midrule
        
        \multicolumn{9}{l}{\textit{\textbf{Qwen2.5-7B-Instruct}}} \\
        Vanilla & 8.82 & 0.840 & 0.790 & 0.510 & 0.835 & 0.987 & 0.928 & 0.254 \\
        LoRA     & 8.39 & 0.840 & 0.790 & 0.507 & 0.833 & 0.988 & 0.925 & 0.258 \\
        SafeDecoding  & 8.32 & \textbf{0.838} & 0.802(\up) & 0.509 & 0.831 & 0.987 & 0.870 & 0.239 \\
        LED       & 7.87 & 0.819 & \textbf{0.808}(\up) & 0.499 & 0.807 & 0.987 & 0.903 & 0.254 \\
        Circuit Breakers  & 7.91 & 0.797 & 0.788 & 0.504 & 0.813 & 0.987 & 0.896 & 0.254 \\
        EVA (text)    & \textbf{8.48} & 0.832 & 0.798(\up) & \textbf{0.510} & \textbf{0.833} & \textbf{0.988}(\up) & \textbf{0.910} & \textbf{0.256}(\up) \\
        \bottomrule
        \end{tabular}
    }
\end{minipage}
\end{table*}

\subsection{Experimental Setup}
\label{subsec:setup}
\partitle{Datasets and metrics} 
We conduct comprehensive evaluations on both LLMs and VLMs. For LLMs, we utilize HarmBench \cite{mazeika2024harmbench} for the editing dataset and evaluate defense performance on HarmBench, AdvBench \cite{zou2023universal}, JailbreakBench \cite{chao2024jailbreakbench}, and MaliciousInstruct \cite{huang2024catastrophic}. Utility is assessed using MT-Bench \cite{zheng2023judging} and seven downstream tasks: \textit{Closed-domain QA} on BoolQ \cite{clark2019boolq} (EM \cite{rajpurkar2016squad}), \textit{Dialogue} on MuTual \cite{cui2020mutual} (Acc), \textit{NER} on CoNLL03 \cite{sang2003introduction} (F1 score\cite{rajpurkar2016squad}), \textit{NLI} on RTE \cite{dagan2005pascal} (Acc), \textit{Reasoning} on GSM8K \cite{cobbe2021training} (Acc), \textit{Sentiment Analysis} on SST2 \cite{socher2013recursive} (Acc), and \textit{Summarization} on SAMSum \cite{gliwa2019samsum} (ROUGE \cite{chin2004rouge}).
For VLMs, we construct a specific dataset for the editing by generating images relevant to Harmbench queries using Stable Diffusion \cite{rombach2022high}, concatenated with OCR images containing harmful tokens. To evaluate defense performance, we employ MM-SafetyBench \cite{liu2024mm} and MultiTrust \cite{zhang2024multitrust}, alongside visual adversarial images generated by \textit{FigStep} \cite{gong2025figstep}, \textit{Hades} \cite{Li-HADES-2024}, and optimization-based attacks \cite{qi2024visual} on the four aforementioned text datasets. VLM utility is evaluated on MM-Vet-v2 \cite{yu2024mm}, MMMU \cite{yue2024mmmu}, and MMStar \cite{chen2024we}.
We report standard performance metrics specific to each utility task, where higher scores indicate better utility. For safety tasks, we report the Attack Success Rate (ASR), where lower indicate better. We employ the HarmBench classifier to detect harmful content. For a dataset $\mathcal{D}_{harm}$ containing harmful queries $\mathbf{q}$, ASR is formally defined as:
\begin{equation}
\text{ASR}(\mathcal{D}_{harm}) = \frac{1}{|\mathcal{D}_{harm}|} \sum{\mathbf{q}\in\mathcal{D}_{harm}} \mathbb{I}(f(\mathbf{q}))
\end{equation}
where $\mathbb{I}$ is the indicator function that returns 1 for successful attacks and 0 otherwise.

\partitle{Models and attacks}
For LLMs, we evaluate five models: Llama-2-7B-chat-hf \cite{touvron2023llama}, Vicuna-7B-v1.5 \cite{zheng2023judging}, Mistral-7B-Instruct-v0.2 \cite{jiang2024mistral}, Llama-3.1-8B-Instruct \cite{grattafiori2024llama}, and Qwen2.5-7B-Instruct \cite{qwen2.5}.
To assess robustness, we generate adversarial prompts across all four datasets using three distinct jailbreak strategies: gradient-based \textit{GCG} \cite{zou2023universal}, genetic algorithm-based \textit{AutoDAN} \cite{liuautodan}, and LLM-assisted iterative attack \textit{PAIR} \cite{chao2025jailbreaking}.
For VLMs, our experiments involve LLaVA-v1.5-7B \cite{liu2023visual}, Qwen2.5-VL-7B \cite{bai2025qwen2}, and InternVL3.5-8B \cite{wang2025internvl3}.
We employ visual jailbreak attacks, including \textit{FigStep} \cite{gong2025figstep}, \textit{Hades} \cite{Li-HADES-2024}, and Visual Adversarial Examples (\textit{ADV-16}) \cite{qi2024visual}. For the \textit{ADV-16}, we evaluate the transferability of adversarial examples generated on LLaVA-v1.5-7B to the other models. And we incorporate attack scenarios from benchmarks such as MM-SafetyBench \cite{liu2024mm} and MultiTrust \cite{zhang2024multitrust}.

\partitle{Baselines}
We compare \textit{EVA} against a diverse set of baselines in the LLM and VLM domains. 
For LLMs, we employ safety fine-tuning approaches including \textit{LoRA} \cite{hu2022lora} and \textit{LED} \cite{zhao-etal-2024-defending-large}, alongside inference-time defense mechanisms like \textit{SafeDecoding} \cite{xu2024safedecoding}. We also include \textit{Circuit Breakers} \cite{zou2024improving} as a representation-level defense baseline.
For VLMs, we select \textit{JailGuard} \cite{10.1145/3724393}, \textit{AdaShield} \cite{wang2024adashield}, and \textit{VLGuard} \cite{zong2024safety} as primary baselines. \textit{AdaShield} is trained on the corresponding datasets following its original protocol. Unless otherwise specified, all other baselines for both LLMs and VLMs utilize their default configurations.
To explicitly isolate and verify the contributions of the multimodal components, we include a text-only variant of our method, denoted as \textit{EVA (text)}. This variant serves as a baseline for VLM experiments.
We use a unified, concise refusal \textit{“I'm sorry, I can't assist with that.”} as the optimization target $\mathbf{y}_{safe}$, deferring the ablation study on optimization target to Section \ref{subsec:ablation}.

\begin{table*}[t]
\centering
\caption{Safety evaluation of \textit{EVA} and baselines on different VLMs across multiple benchmarks. The values report the ASR under various attack settings. \textbf{Bold}: best defense performance.}
\label{tab:vlm_asr}
\resizebox{\textwidth}{!}{
\begin{tabular}{lcccccccccccccccccc}
\toprule
\multirow{2}{*}{\textbf{Datasets}} & \multicolumn{3}{c}{\textbf{HarmBench}} & \multicolumn{3}{c}{\textbf{AdvBench}} & \multicolumn{3}{c}{\textbf{JailbreakBench}} & \multicolumn{3}{c}{\textbf{MaliciousInstruct}} & \multicolumn{3}{c}{\textbf{MM-SafetyBench}} & \multicolumn{3}{c}{\textbf{MultiTrust}} \\
\cmidrule(lr){2-4} \cmidrule(lr){5-7} \cmidrule(lr){8-10} \cmidrule(lr){11-13} \cmidrule(lr){14-16} \cmidrule(lr){17-19}
Methods & \makecell[c]{Fig-\\Step} & \makecell[c]{ADV\\-16} & Hades & \makecell[c]{Fig-\\Step} & \makecell[c]{ADV\\-16} & Hades & \makecell[c]{Fig-\\Step} & \makecell[c]{ADV\\-16} & Hades & \makecell[c]{Fig-\\Step} & \makecell[c]{ADV\\-16} & Hades & SD & TYPO & \makecell[c]{SD+\\TYPO} & \makecell[c]{Typo-\\graphic} & \makecell[c]{Multi-\\modal} & \makecell[c]{Cross-\\modal} \\
\midrule

\multicolumn{19}{l}{\textit{\textbf{LLaVA-1.5-7B}}} \\
Vanilla     & 0.33 & 0.71 & 0.38 & 0.43 & 0.39 & 0.44 & 0.39 & 0.61 & 0.54 & 0.68 & 0.83 & 0.26 & 0.13 & 0.12 & 0.37 & 0.26 & 0.53 & 0.88 \\
JailGuard   & 0.25 & 0.56 & 0.26 & 0.32 & 0.30 & 0.30 & 0.31 & 0.47 & 0.35 & 0.55 & 0.62 & 0.16 & 0.11 & 0.09 & 0.29 & 0.15 & 0.39 & 0.42 \\
Adashield-A & 0.01 & \textbf{0.00} & \textbf{0.00} & \textbf{0.00} & \textbf{0.00} & \textbf{0.00} & 0.02 & \textbf{0.00} & \textbf{0.00} & 0.01 & \textbf{0.00} & \textbf{0.00} & \textbf{0.02} & 0.01 & \textbf{0.10} & 0.02 & 0.08 & 0.07 \\
VLGuard     & \textbf{0.00} & \textbf{0.00} & \textbf{0.00} & \textbf{0.00} & \textbf{0.00} & \textbf{0.00} & \textbf{0.00} & \textbf{0.00} & \textbf{0.00} & \textbf{0.00} & \textbf{0.00} & \textbf{0.00} & 0.04 & 0.02 & 0.14 & 0.04 & \textbf{0.06} & 0.07 \\
EVA(text)   & 0.05 & 0.02 & 0.09 & 0.03 & 0.02 & 0.05 & 0.11 & 0.04 & 0.06 & 0.21 & \textbf{0.00} & 0.01 & 0.09 & 0.08 & 0.33 & 0.07 & 0.42 & \textbf{0.00} \\
EVA         & \textbf{0.00} & \textbf{0.00} & \textbf{0.00} & \textbf{0.00} & \textbf{0.00} & 0.01 & \textbf{0.00} & \textbf{0.00} & \textbf{0.00} & \textbf{0.00} & \textbf{0.00} & \textbf{0.00} & 0.03 & \textbf{0.00} & 0.13 & \textbf{0.01} & 0.13 & \textbf{0.00} \\
\midrule

\multicolumn{19}{l}{\textit{\textbf{Qwen2.5-VL-7B}}} \\
Vanilla     & 0.32 & 0.15 & 0.19 & 0.26 & 0.18 & 0.12 & 0.04 & 0.08 & 0.56 & 0.07 & 0.17 & 0.47 & 0.11 & 0.40 & 0.51 & 0.03 & 0.35 & 0.10 \\
JailGuard   & 0.16 & 0.08 & 0.13 & 0.15 & 0.07 & 0.02 & 0.02 & 0.07 & 0.26 & 0.03 & 0.10 & 0.29 & 0.09 & 0.30 & 0.37 & 0.02 & 0.25 & 0.08 \\
Adashield-A & 0.10 & 0.09 & 0.01 & 0.09 & \textbf{0.00} & 0.04 & \textbf{0.00} & \textbf{0.00} & 0.31 & 0.04 & \textbf{0.00} & 0.12 & 0.05 & 0.27 & 0.27 & \textbf{0.00} & 0.06 & 0.05 \\
VLGuard     & \textbf{0.00} & \textbf{0.00} & \textbf{0.00} & \textbf{0.00} & 0.01 & \textbf{0.00} & \textbf{0.00} & \textbf{0.00} & 0.01 & \textbf{0.00} & \textbf{0.00} & \textbf{0.00} & 0.02 & 0.11 & 0.11 & 0.01 & 0.08 & 0.10 \\
EVA(text)   & 0.03 & \textbf{0.00} & 0.01 & 0.01 & \textbf{0.00} & 0.02 & \textbf{0.00} & \textbf{0.00} & 0.02 & 0.01 & \textbf{0.00} & 0.05 & 0.04 & 0.19 & 0.22 & \textbf{0.00} & 0.14 & \textbf{0.00} \\
EVA         & \textbf{0.00} & \textbf{0.00} & \textbf{0.00} & \textbf{0.00} & \textbf{0.00} & \textbf{0.00} & \textbf{0.00} & \textbf{0.00} & \textbf{0.00} & \textbf{0.00} & \textbf{0.00} & \textbf{0.00} & \textbf{0.01} & \textbf{0.00} & \textbf{0.00} & \textbf{0.00} & \textbf{0.00} & \textbf{0.00} \\
\midrule

\multicolumn{19}{l}{\textit{\textbf{InternVL3.5-8B}}} \\
Vanilla     & 0.22 & 0.11 & 0.17 & 0.05 & 0.04 & 0.03 & 0.15 & 0.10 & 0.09 & 0.30 & 0.22 & 0.21 & 0.14 & 0.48 & 0.48 & 0.03 & 0.33 & 0.07 \\
JailGuard   & 0.16 & 0.08 & 0.13 & 0.04 & 0.03 & 0.02 & 0.11 & 0.07 & 0.06 & 0.22 & 0.16 & 0.15 & 0.10 & 0.36 & 0.35 & 0.02 & 0.24 & 0.05 \\
Adashield-A & 0.01 & 0.05 & 0.04 & \textbf{0.00} & \textbf{0.00} & \textbf{0.00} & 0.01 & 0.02 & 0.07 & 0.02 & 0.01 & 0.01 & 0.03 & 0.12 & 0.09 & \textbf{0.00} & 0.05 & 0.01 \\
VLGuard     & \textbf{0.00} & 0.01 & 0.01 & \textbf{0.00} & \textbf{0.00} & \textbf{0.00} & \textbf{0.00} & \textbf{0.00} & 0.02 & \textbf{0.01} & \textbf{0.00} & \textbf{0.00} & \textbf{0.02} & 0.08 & 0.11 & \textbf{0.00} & 0.04 & 0.02 \\
EVA(text)   & 0.02 & 0.03 & 0.03 & \textbf{0.00} & 0.01 & \textbf{0.00} & 0.03 & 0.02 & 0.01 & 0.05 & 0.03 & 0.02 & 0.05 & 0.14 & 0.09 & 0.01 & 0.15 & \textbf{0.00} \\
EVA         & 0.01 & \textbf{0.00} & \textbf{0.00} & \textbf{0.00} & \textbf{0.00} & \textbf{0.00} & 0.01 & \textbf{0.00} & \textbf{0.00} & 0.02 & \textbf{0.00} & \textbf{0.00} & 0.03 & \textbf{0.06} & \textbf{0.03} & \textbf{0.00} & \textbf{0.00} & \textbf{0.00} \\
\bottomrule
\end{tabular}%
}
\end{table*}

\subsection{Critical Layers and Visual Tokens Selection}
\label{subsec:selection}

\begin{figure}[htbp]
    \centering
    \includegraphics[width=\linewidth]{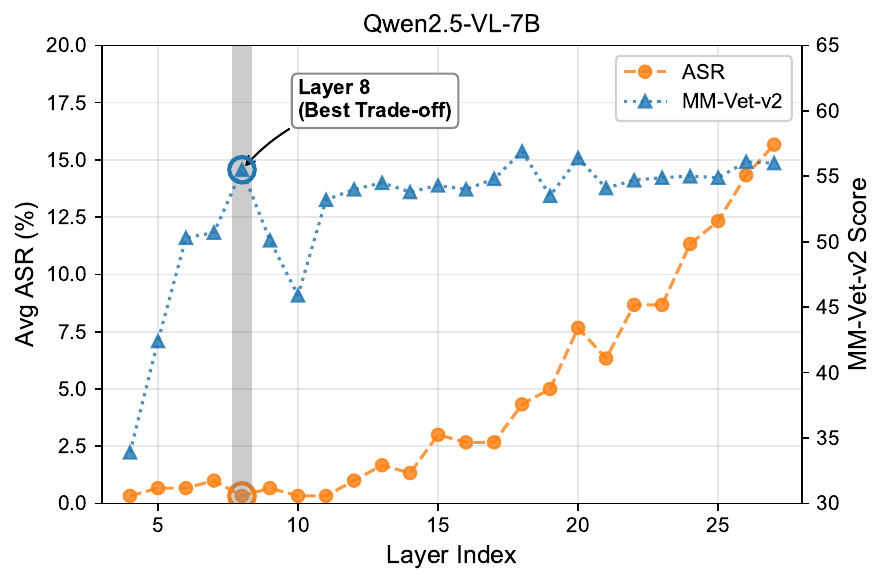}
    \vspace{-2em}
    \caption{Ablation study on Layer selection for Qwen2.5-VL-7B. The left y-axis represents the Avg. ASR(\%). The right y-axis represents the MM-Vet-v2 score. Layer 8 achieves the optimal trade-off.}
    \label{fig:qwen_layer}
\end{figure}

\begin{figure}[htbp]
    \centering
    \includegraphics[width=\linewidth]{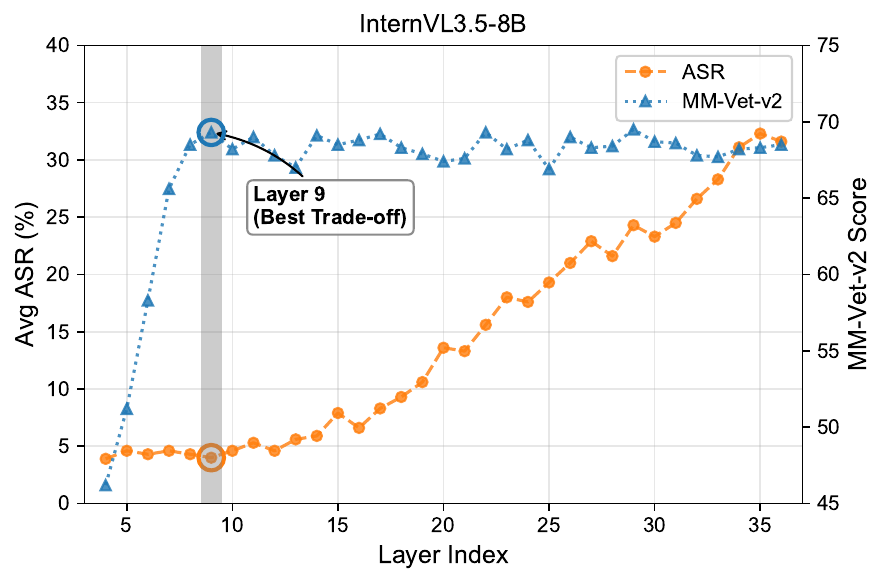}
    \vspace{-2em}
    \caption{Ablation study on Layer selection for InternVL3.5-8B. The left y-axis represents the Avg. ASR(\%). The right y-axis represents the MM-Vet-v2 score. Layer 9 achieves the optimal trade-off.}
    \label{fig:internvl_layer}
    \vspace{-1em}
\end{figure}

\partitle{Critical layers selection}
We target critical layers identified in prior editing works. This choice aligns with recent findings on activation boundaries~\cite{gao2025shaping}, which suggest that jailbreak vulnerabilities are rooted in specific layers. Concretely, we adopt the key layers reported by \textit{MEMIT}~\cite{mengmass} and \textit{EasyEdit}~\cite{wang2024easyedit} for the LLMs. For VLMs, our edits are still applied inside the language model component. In modern VLM architectures, the vast majority of parameters and almost all generative capacity reside in the LLM backbone, while the vision encoder mainly acts as an image feature extractor that feeds embeddings into the language model. This makes it natural to reuse the critical layer configuration of the underlying LLM when performing safety editing on the full VLM.
To verify that this assumption holds in the multimodal setting, we conduct a layer-wise ablation by sweeping the edited layer on VLMs. On Qwen2.5-VL-7B, we vary the edited layer from 4 to 28 (Figure~\ref{fig:qwen_layer}); on InternVL3.5-8B, we vary it from 4 to 36 (Figure~\ref{fig:internvl_layer}). For each layer, we measure the average ASR on MM-SafetyBench and MM-Vet-v2 score.
Our results show that for Qwen2.5-VL-7B, editing at Layer 8 achieves the best safety–utility trade-off, matching the critical layer of the Qwen2.5-7B. Similarly, for InternVL3.5-8B, Layer 9 yields the lowest ASR and highest MM-Vet-v2 score, again coinciding with the layer identified on its LLM backbone Qwen3-8B. These consistent optima across LLMs and VLMs indicate that the editing sweet spot discovered in LLMs largely transfers to VLMs, supporting our choice of directly reusing LLM critical layers when performing safety edits in VLMs.

\partitle{Visual tokens selection}
To determine the most effective strategy for identifying critical visual tokens, we evaluated five selection methods against our attention-based approach. All methods select one image token for editing in each input. We conducted experiments on the Qwen2.5-VL-7B, assessing the ASR on the MM-SafetyBench (SD+TYPO subsets) to measure safety, and the MM-Vet-v2 score to measure general capabilities.
The \textit{Random} method selects an image token arbitrarily from the visual sequence, while the \textit{Fixed} strategy consistently targets the first image token. Moving to embedding-based methods, we evaluate methods based on the interaction between the key vector of image tokens $k_{m_i}$ and the key vector of the harmful text token $k_{t}$. The \textit{L2 Distance} method selects the token minimizing the Euclidean distance \cite{golub2013matrix}, formulated as $i^* = \operatorname*{argmin}_{i} \| k_{m_i} - k_{t} \|_2$. Similarly, the \textit{Cosine Similarity} \cite{salton1975vector} method selects the token with the highest semantic alignment by maximizing the cosine similarity: $i^* = \operatorname*{argmax}_{i} \frac{k_{m_i} \cdot k_{t}}{\|k_{m_i}\| \|k_{t}\|}$.
\begin{table}[htbp]
\centering
\caption{Comparison of different image token selection strategies on Qwen2.5-VL-7B. \textbf{Bold} indicates the best result in each column.}
\resizebox{0.9\linewidth}{!}{
\begin{tabular}{lcc}
\toprule
\textbf{Selection Method} & \textbf{MM-SafetyBench} & \textbf{MM-Vet-v2} \\
                          & ASR $\downarrow$    & Score $\uparrow$ \\
\midrule
Vanilla                    & 0.51            & 56.9 \\
\midrule
Random                     & \textbf{0.00}            & 10.4 \\
Fixed                      & \textbf{0.00}            & 12.1 \\
L2 Distance                & 0.08            & 54.4 \\
Cosine Similarity          & 0.05            & 55.2 \\
Gradient-based             & 0.02            & \textbf{56.0} \\
Attention-based (Ours)     & \textbf{0.00}   & 55.5 \\
\bottomrule
\end{tabular}
}
\label{tab:token_selection_methods}
\end{table}
We further compare our approach against a gradient-based attribution baseline. This method measures the sensitivity of the model's affirmative output (e.g., ``Sure, here is'') to changes in the input image tokens. The most critical token is identified by maximizing the gradient norm with respect to the target loss $\mathcal{L}_{\text{target}}$: $i^* = \operatorname*{argmax}_{i} \left\lVert \nabla_{m_i} \mathcal{L}_{\text{target}} \right\rVert.$
As shown in Table~\ref{tab:token_selection_methods}, naive strategies (\textit{Random}, \textit{Fixed}) cause a catastrophic collapse in general capabilities, with MM-Vet-v2 scores dropping to 10. Embedding-based methods (\textit{L2}, \textit{Cosine}) and \textit{Gradient-based} approach preserves model utility, but they all necessitate additional computational overhead for vector compute or backward propagation. And these methods fail to fully mitigate jailbreak risks (ASR $>0$).
In contrast, our attention-based mechanism utilizes the attention scores assigned by the text tokens to the image tokens. Since these scores are a byproduct of the standard forward pass, our method incurs zero additional computational overhead while maintaining safety and high utility, providing efficient alternative for identifying harmful visual triggers.

\subsection{Safety and Utility Evaluation}
\label{subsec:effectiveness}

\partitle{Safety evaluation} 
We evaluate the defense performance of \textit{EVA} across both LLMs and VLMs. Table \ref{tab:llm_asr} and Table \ref{tab:vlm_asr} present the ASR under various adversarial scenarios.
For LLMs, as shown in Table \ref{tab:llm_asr}, \textit{EVA} demonstrates superior robustness compared to all baselines. On Vicuna-7B-v1.5, \textit{EVA} significantly reduces the ASR of \textit{AutoDAN} attacks from 0.69 to 0.04, outperforming \textit{LED} (0.11) and \textit{Circuit Breakers} (0.10). Notably, on stronger models like Llama3.1-8B-Instruct, and Qwen2.5-7B-Instruct, \textit{EVA} achieves near-perfect defense with ASRs consistently dropping to 0 across four datasets. In contrast, while \textit{LED} shows decent defense in some specific settings, it exhibits instability (e.g., 0.34 ASR on JailbreakBench for Vicuna), and \textit{SafeDecoding} often fails to suppress attacks effectively on Mistral and Qwen.
For VLMs, the results in Table \ref{tab:vlm_asr} confirm that \textit{EVA} effectively generalizes to the multimodal domain. On LLaVA-1.5-7B, \textit{EVA} drastically reduces the ASR of the \textit{ADV-16} attack from 0.71 to 0.00, and suppresses \textit{Crossmodal} attacks in MultiTrust from 0.88 to 0.00. This significantly outperforms \textit{JailGuard}, which retains a high ASR of 0.56 on \textit{ADV-16}. Similarly, on Qwen2.5-VL-7B and InternVL3.5-8B, \textit{EVA} consistently achieves 0.00 ASR against sophisticated visual jailbreaks like \textit{Hades}. Furthermore, on MM-SafetyBench (SD+TYPO), where Qwen2.5-VL-7B reaches an ASR of 0.51, \textit{EVA} completely mitigates the threat (0.00), demonstrating superior reliability compared to baselines like \textit{AdaShield} and \textit{VLGuard} across diverse datasets.

\begin{table}[t]
\centering
\caption{Utility evaluation of \textit{EVA} and baselines on three general multimodal benchmarks. \textbf{Bold}: best performance.}
\label{tab:vlm_utility}
\resizebox{0.7\linewidth}{!}{%
\begin{tabular}{lccc}
\toprule
\textbf{Methods} & \textbf{MM-Vet-v2} & \textbf{MMMU} & \textbf{MMStar} \\
\midrule

\multicolumn{4}{l}{\textit{\textbf{LLaVA-1.5-7B}}} \\
Vanilla     & 28.5 & 31.3 & 32.5 \\
JailGuard   & 26.5 & 29.8 & 31.0 \\
Adashield-A & 14.1 & 19.7 & 19.2 \\
VLGuard     & 28.1 & 32.1 & \textbf{32.3} \\
EVA(text)   & \textbf{28.6} & 33.3 & 27.3 \\
EVA         & 28.3 & \textbf{33.4} & \textbf{32.3} \\
\midrule

\multicolumn{4}{l}{\textit{\textbf{Qwen2.5-VL-7B}}} \\
Vanilla     & 56.9 & 49.9 & 60.1 \\
JailGuard   & 54.5 & 47.5 & 59.2 \\
Adashield-A & 33.2 & 35.4 & 39.8 \\
VLGuard     & 46.6 & \textbf{50.2} & \textbf{61.6} \\
EVA(text)   & \textbf{55.7} & 49.6 & 61.1 \\
EVA         & 55.5 & 48.2 & 60.9 \\
\midrule

\multicolumn{4}{l}{\textit{\textbf{InternVL3.5-8B}}} \\
Vanilla     & 69.1 & 57.6 & 68.3 \\
JailGuard   & 64.4 & 55.8 & 66.5 \\
Adashield-A & 46.2 & 38.4 & 44.7 \\
VLGuard     & 67.5 & 57.1 & 67.0 \\
EVA(text)   & 69.0 & 57.0 & \textbf{67.6} \\
EVA         & \textbf{69.3} & \textbf{57.6} & 67.5 \\
\bottomrule
\end{tabular}%
}
\vspace{-1em}
\end{table}

\partitle{Utility evaluation} 
We report the utility scores for LLMs and VLMs in Table \ref{tab:llm_utility} and Table \ref{tab:vlm_utility}, respectively.
For LLMs, on Vicuna-7B-v1.5 and Llama3.1-8B-Instruct, \textit{EVA (text)} achieves MT-Bench scores of 6.84 and 7.44, respectively, surpassing the \textit{LED} (3.70 and 7.29) and even outperforming the \textit{Vanilla} model in several cases. Specifically, \textit{EVA (text)} shows improvements in complex tasks such as \textit{Reasoning}, \textit{NLI}, and \textit{Summarization}. This contrasts with \textit{SafeDecoding} and \textit{LED}, which suffer from utility degradation.
For VLMs, \textit{EVA} demonstrates stability in maintaining multimodal reasoning capabilities. On benchmarks like MM-Vet-v2, MMMU, and MMStar, \textit{EVA} achieves scores comparable to the \textit{Vanilla} model (e.g., 69.3 vs. 69.1 on MM-Vet-v2 for InternVL3.5). Conversely, other defense methods such as \textit{JailGuard} and \textit{AdaShield} exhibit severe performance collapses on several benchmarks, suggesting they may compromise the model's gerenal abilities. In summary, \textit{EVA} successfully establishes a robust safety shield while incurring minimal to no utility loss, achieving the best trade-off among all baselines.
\begin{tcolorbox}[
    colback=gray!8,
    colframe=black!60,
    boxrule=0.6pt,
    arc=1.5mm,
    title=\textbf{Summary of Findings},
    left=1.2mm,
    right=1.2mm,
    top=0.8mm,
    bottom=0.8mm
]
\small
\begin{itemize}[leftmargin=1.2em,itemsep=2pt,topsep=2pt]
    \item \textbf{Strong safety.} \textit{EVA} achieves the strongest or near-strongest defense performance on both LLMs and VLMs, reducing ASR from 69\% to 4\% on Vicuna and to 0\% on stronger models such as Llama3.1 and Qwen2.5.
    \item \textbf{Low utility loss.} \textit{EVA} largely preserves utility and often remains close to, or even exceeds, the \textit{Vanilla} model, e.g., 69.3 vs. 69.1 on MM-Vet-v2 for InternVL and 33.4 on MMMU for LLaVA.
    \item \textbf{Best trade-off.} Compared with prior defenses, \textit{EVA} provides the best overall balance between safety and utility, achieving 0\% ASR on MM-SafetyBench for Qwen2.5-VL while retaining strong utility scores of 55.5/48.2/60.9 on MM-Vet-v2/MMMU/MMStar.
\end{itemize}
\end{tcolorbox}

\subsection{Adaptive Attacks Evaluation}
\label{subsec:adaptive_attacks}

\partitle{Adaptive attacks on LLMs}
Adaptive attacks \cite{tramer2020adaptive} provide a stronger test of robustness, but \textit{EVA} remains effective. To evaluate this setting, we regenerate adversarial test cases directly on each edited LLM using \textit{GCG}, \textit{AutoDAN}, and \textit{PAIR} across four datasets, and report the resulting ASR. The hyperparameter settings for adaptive attack generation are kept identical to those used when generating adversarial test cases on the corresponding vanilla model.
As shown in Table~\ref{tab:adaptive_llm_asr}, adaptive attacks increase the ASR of \textit{EVA (text)} on several models, confirming that they are stronger than the standard setting. However, the increase remains limited relative to the corresponding \textit{Vanilla} models. For example, on Vicuna-7B, the average ASR rises from 0.068 to 0.112, but remains far below the vanilla ASR of 0.821. Similar trends are observed on Mistral-7B (0.122 vs. 0.089, compared with 0.863 for vanilla) and Qwen2.5-7B (0.113 vs. 0.012, compared with 0.634 for vanilla). On Llama2-7B and Llama3.1-8B, adaptive attacks remain largely ineffective, with average ASR staying at 0.000 and 0.005. Overall, although adaptive attacks can partially recover attack success on some models, \textit{EVA} still preserves strong robustness.

\partitle{Adaptive attacks on VLMs}
For VLMs, we conduct adaptive \textit{ADV-16} evaluation on LLaVA-1.5-7B, where \textit{EVA} also remains effective. Most attack settings considered in this work, including \textit{FigStep}, \textit{Hades}, MM-SafetyBench, and MultiTrust, are dataset or input-construction-based and do not depend on the target model during attack generation. Hence, they do not have a separate adaptive regeneration setting on the edited model. Among our evaluated VLM attacks, only \textit{ADV-16} depends on the target model, since it requires gradients to be back-propagated to the input image. In our experiments, only LLaVA-1.5-7B supports this gradient-based image optimization reliably.
As shown in Table~\ref{tab:adaptive_vlm_asr}, adaptive visual attacks partially recover attack success, increasing the average ASR from 0.000 to 0.125. This remains far below the vanilla ASR of 0.635, further showing that \textit{EVA} remains effective even under adaptive gradient-based visual attacks.

\begin{table}[t]
\centering
\scriptsize
\setlength{\tabcolsep}{1.8pt}
\caption{Adaptive attacks evaluation on LLMs. We report ASR under \textit{GCG}, \textit{AutoDAN}, and \textit{PAIR} across four datasets, with the last column showing the average ASR. Lower is better.}
\label{tab:adaptive_llm_asr}
\resizebox{\linewidth}{!}{
\begin{tabular}{lccccccccccccc}
\toprule
\textbf{Datasets} & \multicolumn{3}{c}{\textbf{HarmBench}} & \multicolumn{3}{c}{\textbf{AdvBench}} & \multicolumn{3}{c}{\textbf{JailbreakBench}} & \multicolumn{3}{c}{\textbf{MaliciousInstruct}} & \textbf{Avg.} \\
\cmidrule(lr){2-4} \cmidrule(lr){5-7} \cmidrule(lr){8-10} \cmidrule(lr){11-13}
\textbf{Methods} & GCG & \makecell[c]{Auto\\DAN} & PAIR & GCG & \makecell[c]{Auto\\DAN} & PAIR & GCG & \makecell[c]{Auto\\DAN} & PAIR & GCG & \makecell[c]{Auto\\DAN} & PAIR & \\
\midrule

\multicolumn{14}{l}{\textit{\textbf{Vicuna-7B-v1.5}}} \\
Vanilla & 0.92 & 0.69 & 0.80 & 0.89 & 0.78 & 0.75 & 0.89 & 0.73 & 0.77 & 0.94 & 0.83 & 0.86 & 0.821 \\
EVA (text) & 0.11 & 0.04 & 0.10 & 0.02 & 0.02 & 0.05 & 0.17 & 0.08 & 0.11 & 0.01 & 0.05 & 0.05 & 0.068 \\
\quad + Adaptive attacks & 0.17 & 0.06 & 0.08 & 0.11 & 0.04 & 0.06 & 0.21 & 0.10 & 0.15 & 0.13 & 0.09 & 0.14 & 0.112 \\
\midrule

\multicolumn{14}{l}{\textit{\textbf{Llama2-7B-chat}}} \\
Vanilla & 0.42 & 0.23 & 0.02 & 0.39 & 0.19 & 0.01 & 0.46 & 0.27 & 0.04 & 0.45 & 0.30 & 0.00 & 0.232 \\
EVA (text) & 0.00 & 0.00 & 0.00 & 0.00 & 0.00 & 0.00 & 0.00 & 0.00 & 0.00 & 0.01 & 0.00 & 0.00 & 0.001 \\
\quad + Adaptive attacks & 0.00 & 0.00 & 0.00 & 0.00 & 0.00 & 0.00 & 0.00 & 0.00 & 0.00 & 0.00 & 0.00 & 0.00 & 0.000 \\
\midrule

\multicolumn{14}{l}{\textit{\textbf{Mistral-7B-Instruct-v0.2}}} \\
Vanilla & 0.80 & 0.87 & 0.88 & 0.56 & 0.97 & 0.82 & 0.79 & 0.95 & 0.89 & 0.94 & 0.96 & 0.92 & 0.863 \\
EVA (text) & 0.06 & 0.08 & 0.15 & 0.02 & 0.11 & 0.00 & 0.02 & 0.11 & 0.16 & 0.14 & 0.12 & 0.10 & 0.089 \\
\quad + Adaptive attacks & 0.09 & 0.12 & 0.14 & 0.07 & 0.17 & 0.02 & 0.06 & 0.19 & 0.14 & 0.03 & 0.28 & 0.15 & 0.122 \\
\midrule

\multicolumn{14}{l}{\textit{\textbf{Llama3.1-8B-Instruct}}} \\
Vanilla & 0.56 & 0.20 & 0.12 & 0.47 & 0.31 & 0.04 & 0.46 & 0.31 & 0.11 & 0.69 & 0.46 & 0.04 & 0.314 \\
EVA (text) & 0.00 & 0.00 & 0.00 & 0.01 & 0.00 & 0.00 & 0.02 & 0.00 & 0.00 & 0.00 & 0.00 & 0.00 & 0.003 \\
\quad + Adaptive attacks & 0.00 & 0.00 & 0.01 & 0.00 & 0.00 & 0.02 & 0.00 & 0.00 & 0.00 & 0.00 & 0.00 & 0.03 & 0.005 \\
\midrule

\multicolumn{14}{l}{\textit{\textbf{Qwen2.5-7B-Instruct}}} \\
Vanilla & 0.49 & 0.74 & 0.66 & 0.46 & 0.80 & 0.39 & 0.42 & 0.80 & 0.55 & 0.80 & 0.96 & 0.54 & 0.634 \\
EVA (text) & 0.00 & 0.00 & 0.03 & 0.00 & 0.00 & 0.01 & 0.00 & 0.01 & 0.05 & 0.00 & 0.00 & 0.04 & 0.012 \\
\quad + Adaptive attacks & 0.12 & 0.15 & 0.06 & 0.07 & 0.09 & 0.02 & 0.11 & 0.13 & 0.07 & 0.22 & 0.23 & 0.08 & 0.113 \\
\bottomrule
\end{tabular}
}
\end{table}

\begin{table}[t]
\centering
\caption{Adaptive attacks evaluation on LLaVA-1.5-7B. We report ASR under \textit{ADV-16} across four datasets, with the last column showing the average ASR. Lower is better.}
\label{tab:adaptive_vlm_asr}
\resizebox{\linewidth}{!}{
\begin{tabular}{lccccc}
\toprule
\textbf{Methods} & \makecell[c]{\textbf{Harm}\\\textbf{Bench}} & \makecell[c]{\textbf{Adv}\\\textbf{Bench}} & \makecell[c]{\textbf{Jailbreak}\\\textbf{Bench}} & \makecell[c]{\textbf{Malicious}\\\textbf{Instruct}} & \textbf{Avg.} \\
\midrule
Vanilla & 0.71 & 0.39 & 0.61 & 0.83 & 0.635 \\
EVA & 0.00 & 0.00 & 0.00 & 0.00 & 0.000 \\
\quad + Adaptive attacks & 0.12 & 0.08 & 0.13 & 0.17 & 0.125 \\
\bottomrule
\end{tabular}
}
\end{table}

\subsection{Generalization and Interpretability}
\label{subsec:generalization_interpretability}

In this section, we investigate the transferability of \textit{EVA} across different harmful behaviors and provide an interpretability analysis to explain the underlying mechanism.

\begin{figure}[h]
    \centering
    \includegraphics[width=0.9\linewidth]{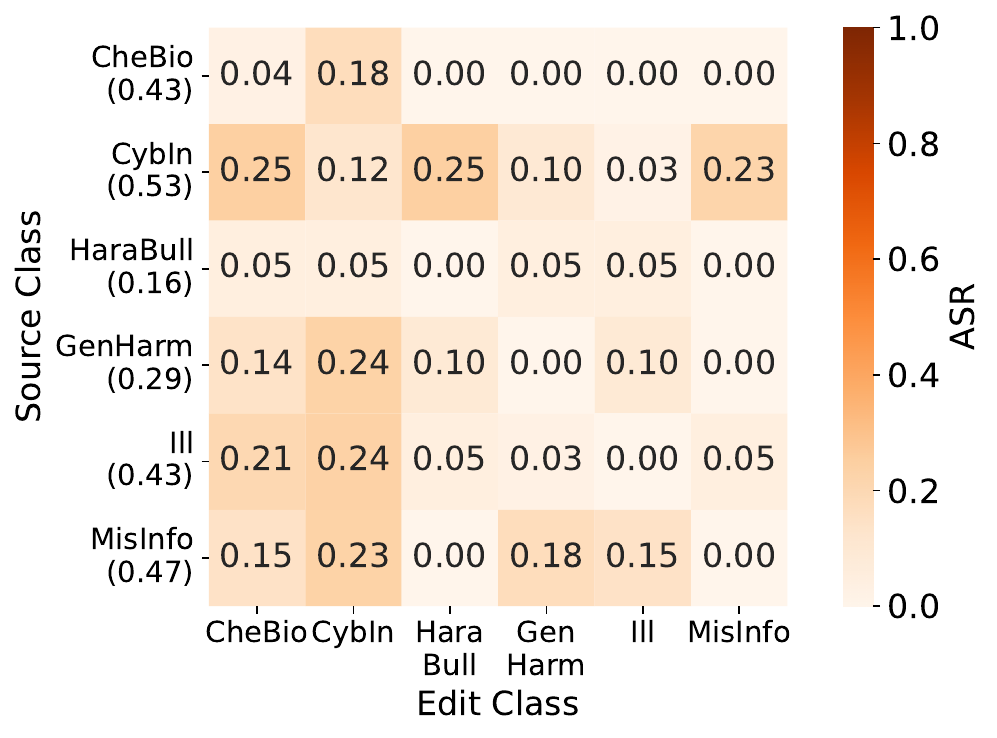}
    \vspace{-1em}
    \caption{ASR heatmap on Llama2-7B against GCG attacks evaluated on the HarmBench dataset. Original ASR in parentheses.}
    \label{fig:llm_heatmap}
\end{figure}

\begin{figure}[h]
    \vspace{-1em}
    \centering
    \includegraphics[width=0.9\linewidth]{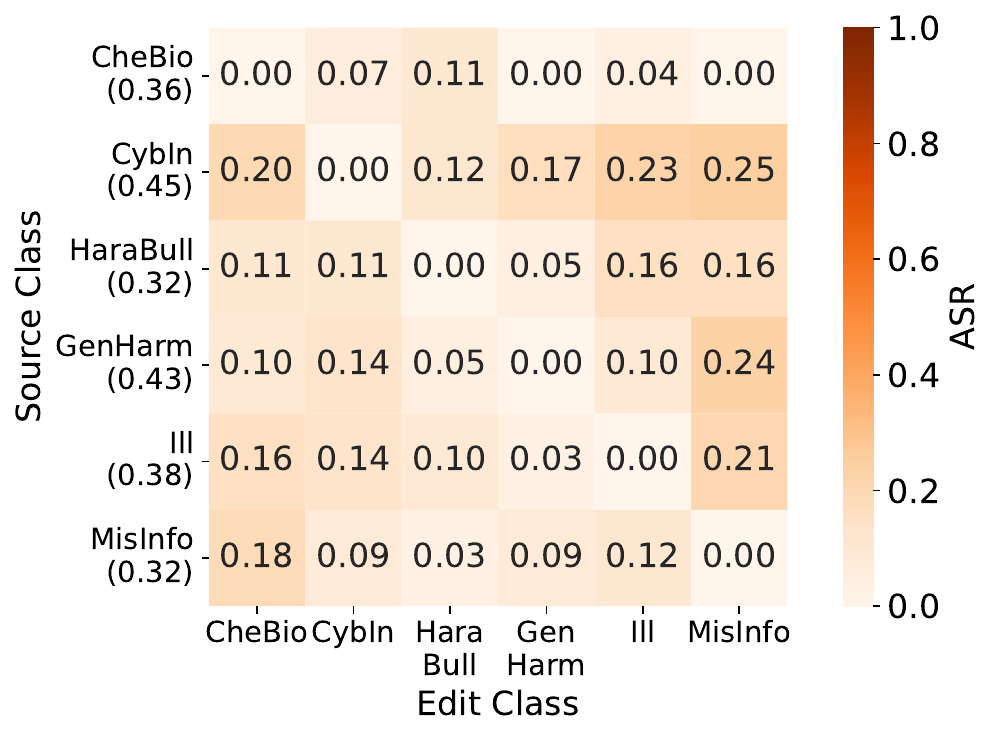}
    \vspace{-1em}
    \caption{ASR heatmap on LLaVA-1.5-7B against Hades attacks evaluated on the HarmBench dataset. Original ASR in parentheses.}
    \label{fig:vlm_heatmap}
\end{figure}

\partitle{Cross-behavior generalization} 
We evaluate whether editing a model on a single harmful category generalizes to defend against unseen categories. Our experiments utilize the six categories defined in the HarmBench dataset: Chemical and Biological (CheBio), Cybercrime and Intrusion (CybIn), Harassment and Bullying (HaraBull), General Harmful (GenHarm), Illegal Activities (Ill), and Misinformation (MisInfo). We conduct single category edits on Llama2-7B-chat (against \textit{GCG}) and LLaVA-1.5-7B (against \textit{Hades}), presenting ASR heatmaps where off-diagonal elements represent defense performance on unedited categories. As shown in Figure \ref{fig:llm_heatmap} and \ref{fig:vlm_heatmap}, \textit{EVA} exhibits transferability across behavior categories. For instance, editing solely on a category effectively mitigates threats in other domains, dropping the ASR. This indicates that \textit{EVA} captures common features of attacks and recognizes underlying malicious intent, rather than merely memorizing specific refusal templates.

\begin{figure}[h]
    \centering
    \subfloat[Llama2: $\mathbf{k}$ of text tokens across behaviors.\label{fig:pca1}]{
        \includegraphics[width=0.23\textwidth]{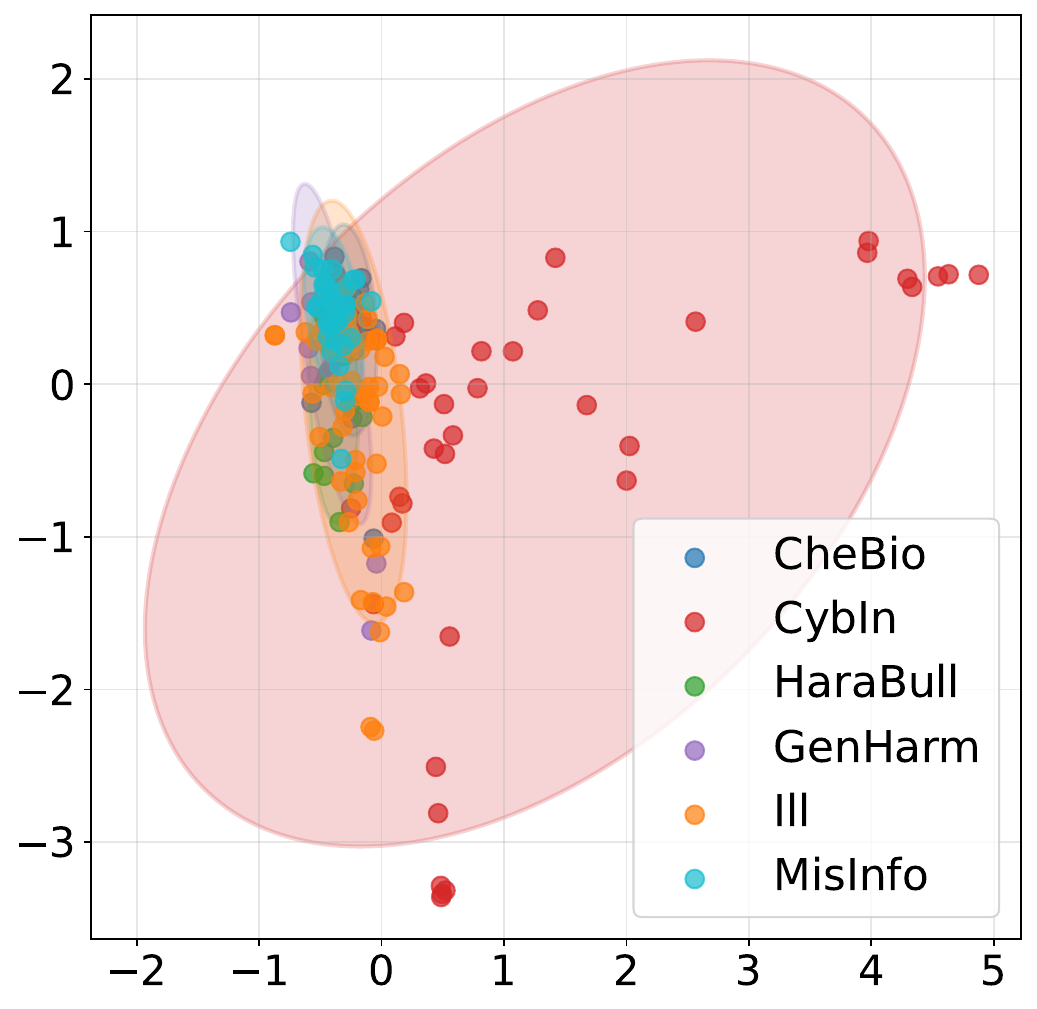}
    }
    \hfill
    \subfloat[Llama2: $\mathbf{k}$ of text tokens across datasets.\label{fig:pca2}]{
        \includegraphics[width=0.23\textwidth]{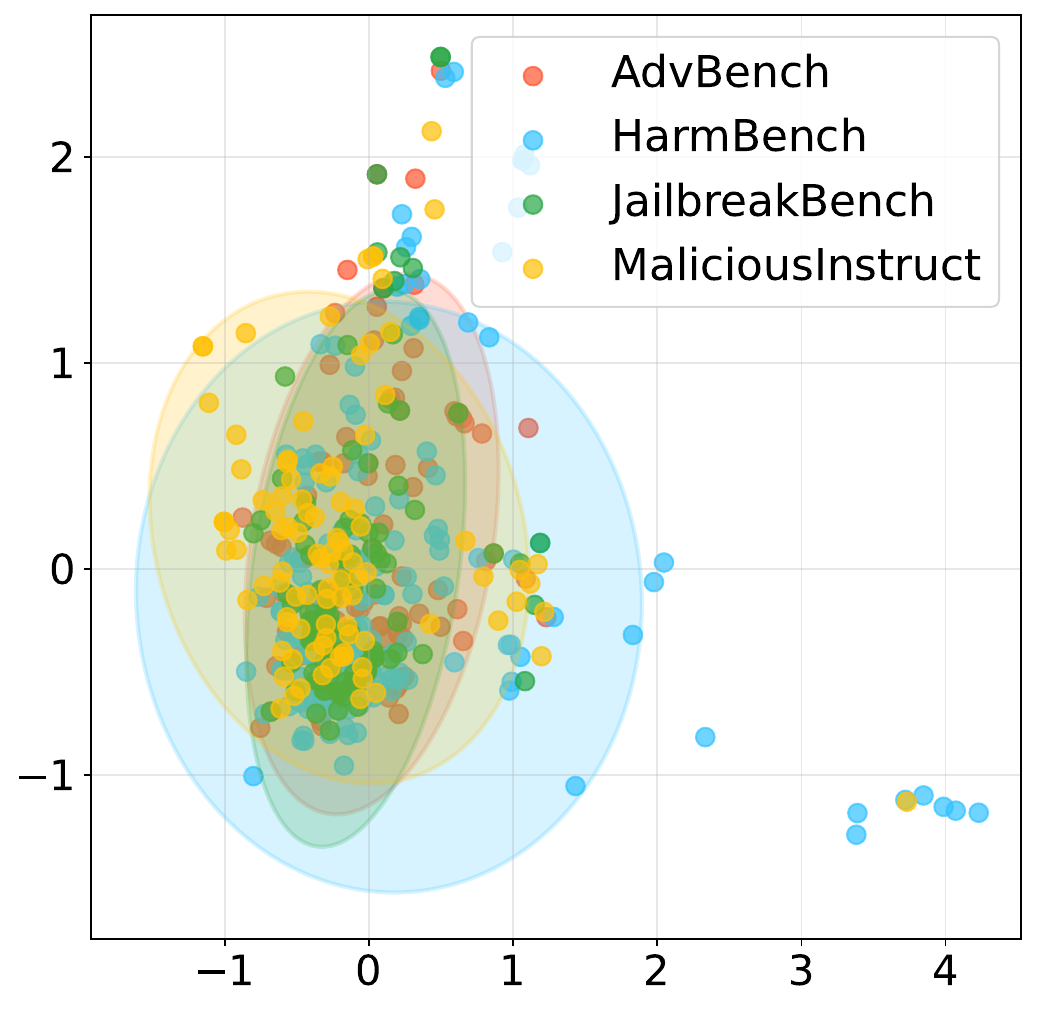}
    }
    
    \subfloat[Qwen2.5-VL: $\mathbf{k}$ of image tokens across behaviors.\label{fig:pca3}]{
        \includegraphics[width=0.23\textwidth]{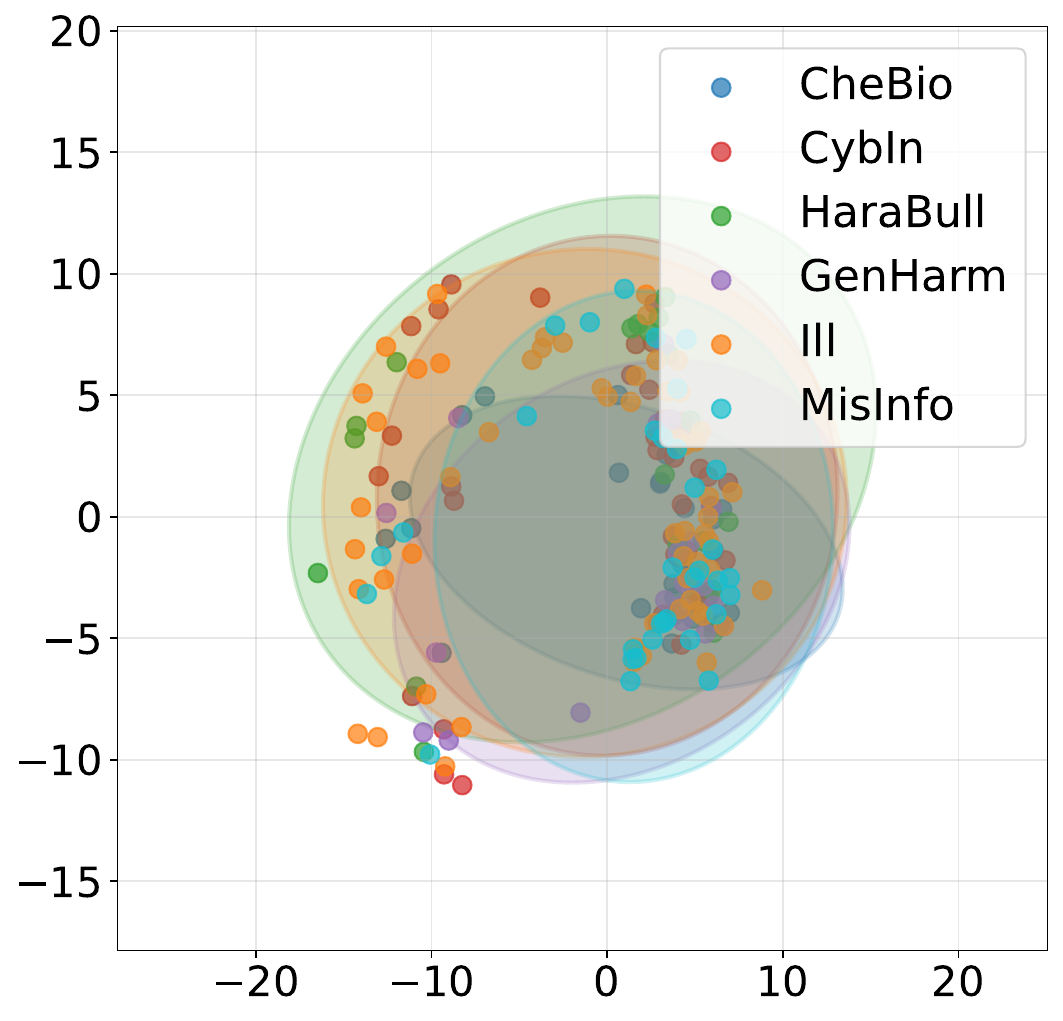}
    }
    \hfill
    \subfloat[Qwen2.5-VL: $\mathbf{k}$ of image tokens across datasets.\label{fig:pca4}]{
        \includegraphics[width=0.23\textwidth]{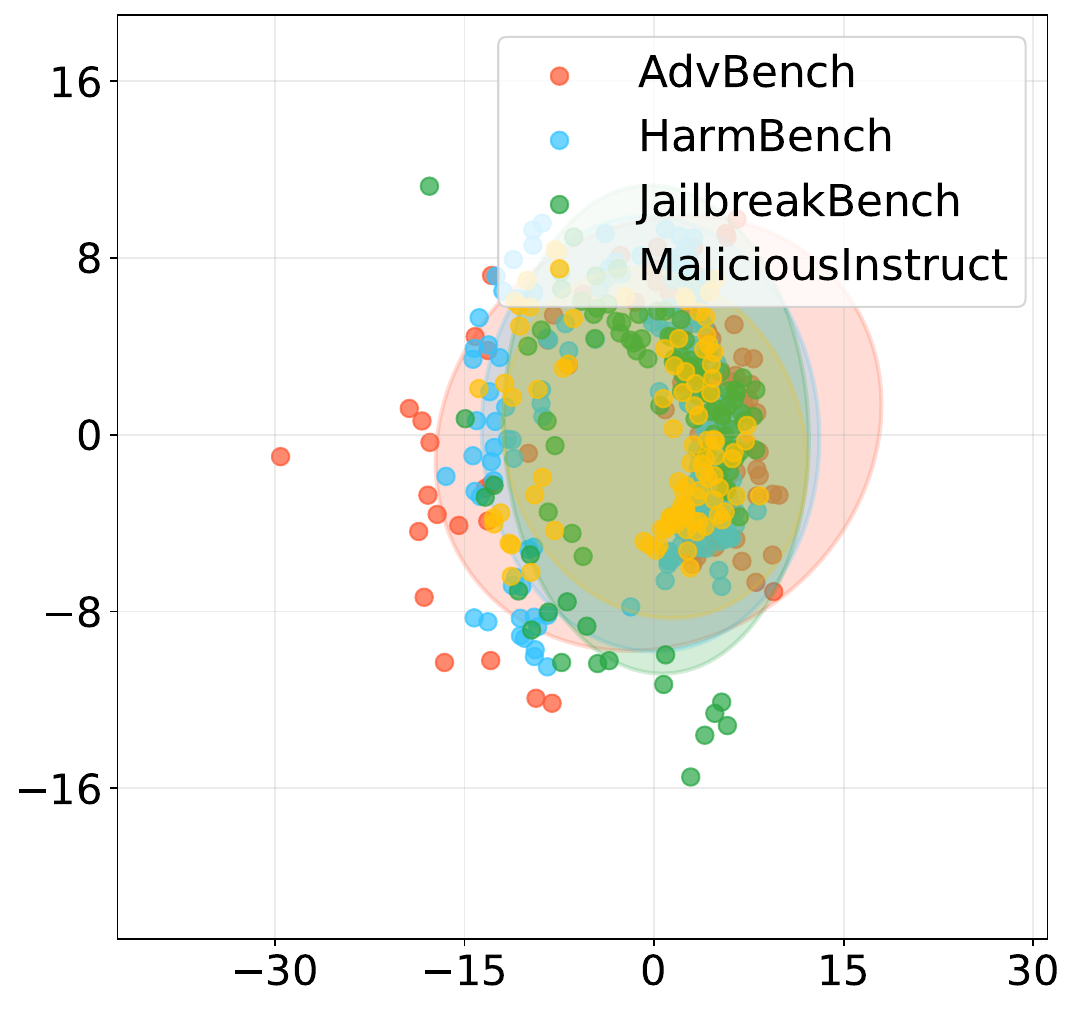}
    }
    
    \caption{PCA visualizations of $\mathbf{k}$ at the target layer $l^*$. \textbf{Top Row:} Text token representations in Llama2. \textbf{Bottom Row:} Image token representations in Qwen2.5-VL.}
    \label{fig:two_pca}
\end{figure}

\partitle{Interpretability via representation analysis}
To understand why \textit{EVA} attains defense generalization with limited editing data, we analyze the internal representations associated with harmful inputs. Concretely, we extract the key vectors $\mathbf{k}$ at the critical layer $l^*$ and visualize their distribution using Principal Component Analysis (PCA) \cite{PCA}. We conduct this analysis in two settings: (i) harmful text tokens in Llama2-7B-chat and (ii) selected image tokens in Qwen2.5-VL-7B-Instruct.
As shown in Figure~\ref{fig:two_pca}, $\mathbf{k}$ vectors corresponding to harmful inputs exhibit substantial overlap and form a coherent cluster in the projected space. For Llama2, keys from different harmful behaviors as well as different datasets concentrate in a similar region, suggesting a shared representation structure beyond dataset-specific artifacts. Notably, the same phenomenon emerges in the visual modality. For Qwen2.5-VL, keys of harmful image tokens also occupy a common subspace across behaviors and data sources.
These observations indicate that the models encode the underlying malicious intent in a consistent across datasets manner at layer $l^*$. Therefore, editing the shared features within this cluster allows \textit{EVA} to correct the model's responses for a broad family of harmful inputs, explaining its transferability to unseen categories and datasets.

\subsection{Decoupling Effects and Contextual Synergy in Visual Editing}
\label{subsec:ocr_case_study}
A question regarding \textit{EVA} is its mechanism of action: Does \textit{EVA} simply suppress all visual information or precisely target specific harmful semantics? To answer this, we design a controlled experiment to decouple visual imagery from optical character recognition (OCR) triggers.

\begin{figure}[htbp]
    \centering
    \includegraphics[width=\linewidth]{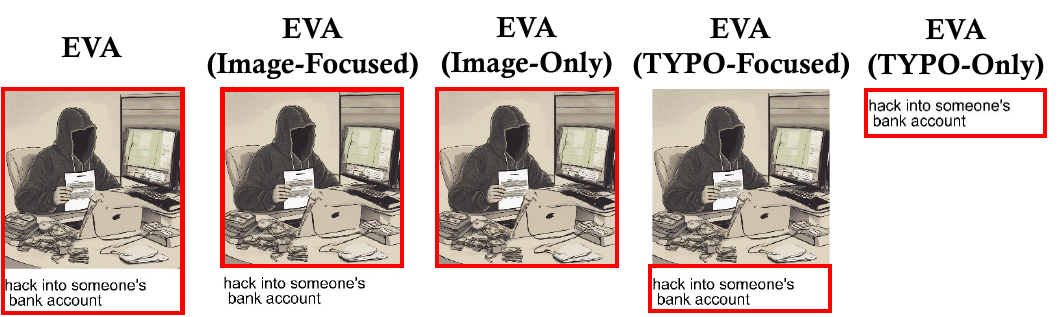}
    \caption{Illustration of \textit{EVA} variations. The red bounding boxes indicate the regions where the model is permitted to select tokens for editing.}
    \label{fig:eva_v}
\end{figure}

\partitle{Experimental setup}
We utilize the MM-SafetyBench, focusing on three attack scenarios: purely visual harmful image generated by Stable Diffusion (SD), text-based OCR attacks (TYPO), and composite samples containing both harmful image and OCR (SD+TYPO). 
We compare four variations of \textit{EVA} to analyze how the selection of editing tokens impacts defense generalization, as illustrated in Figure~\ref{fig:eva_v}, \textit{EVA (Image-Focused)}, which uses composite images but restricts selection to non-OCR visual regions; \textit{EVA (Image-Only)}, which utilizes purely visual harmful images without OCR; \textit{EVA (TYPO-Focused)}, which uses composite images but limits token selection to OCR regions; and \textit{EVA (TYPO-Only)}, which employs images containing only OCR regions.

\partitle{Results and analysis}
Table~\ref{tab:ocr_granularity} presents the ASR, from which we identify two key properties of \textit{EVA}'s defense mechanism.
First, \textit{EVA} demonstrates semantic precision by targeting the specific source of harm. We observe a clear decoupling effect where edits applied to visual regions (\textit{Image-Focused/Only}) primarily mitigate visual attacks (SD), while edits applied to text regions (\textit{TYPO-Focused/Only}) effectively neutralize OCR attacks (TYPO). This confirms that \textit{EVA} operates with granularity rather than suppresses global visual features. It successfully identifies and neutralizes specific triggers while leaving unrelated representations largely unaffected.
Second, we observe contextual synergy, where editing in a rich context improves generalization. A finding is that \textit{Focused} variants outperform their \textit{Only} counterparts. For instance, \textit{Image-Focused} generalizes better to OCR attacks than \textit{Image-Only}, even though both strictly edit visual tokens. The difference lies in the context; in the \textit{Focused} setting, the model observes both harmful image and harmful OCR simultaneously. Even if we only edit the visual tokens, the calculation is conditioned on the presence of the nearby harmful OCR. This suggests that \textit{EVA} captures correlations. By editing a token within a composite toxic context, the defense learns a more robust safety boundary that accounts for the interplay between image and OCR inputs rather than treating them in isolation.
\begin{table}[htbp]
\centering
\caption{The ASR across different attack types when applying variations of \textit{EVA}.}
\label{tab:ocr_granularity}
\resizebox{\linewidth}{!}{
\begin{tabular}{l|ccc|ccc}
\toprule
\textbf{Model}& \multicolumn{3}{c|}{\textbf{Qwen2.5-VL-7B}} & \multicolumn{3}{c}{\textbf{InternVL3.5-8B}} \\
Scenario            & SD   & TYPO & SD+TYPO & SD   & TYPO & SD+TYPO \\
\midrule
Vanilla             & 0.11 & 0.40 & 0.51    & 0.14 & 0.48 & 0.48    \\
\midrule
EVA                 & 0.01 & \textbf{0.00} & \textbf{0.00}    & \textbf{0.03} & 0.06 & \textbf{0.03} \\
EVA (Image-Focused) & 0.01 & 0.15 & 0.12    & 0.06 & 0.17 & 0.07 \\
EVA (Image-Only)    & \textbf{0.00} & 0.18 & 0.13    & \textbf{0.03} & 0.36 & 0.16 \\
EVA (TYPO-Focused)  & 0.05 & \textbf{0.00} & 0.25    & 0.10 & 0.04 & 0.27 \\
EVA (TYPO-Only)     & 0.10 & \textbf{0.00} & 0.40    & 0.13 & \textbf{0.02} & 0.39 \\
\bottomrule
\end{tabular}
}
\end{table}
These results demonstrate that \textit{EVA} is both precise and context-aware. It does not simply penalize all visual inputs; rather, it learns to neutralize specific harmful semantics based on the multimodal context.

\begin{table}[htbp]
\centering
\caption{Efficiency comparison of defense methods on Vicuna-7B-v1.5 and InternVL3.5-8B. \textbf{Bold} indicates the best performance.}
\label{tab:efficiency}
\resizebox{\linewidth}{!}{
\begin{tabular}{llccc}
\toprule
\textbf{Base Model} & \textbf{Method} &
\makecell{\textbf{Training}\\\textbf{Time}$\downarrow$} &
\makecell{\textbf{Inference}\\\textbf{Overhead}$\downarrow$} &
\textbf{Avg. ASR (\%) $\downarrow$} \\
\midrule
\multirow{5}{*}{Vicuna-7B-v1.5} 
& LoRA & 1.5 hr & \textbf{1$\times$} & 23.2\% \\
& SafeDecoding & 0.6 hr & 1.07$\times$ & 10.7\% \\
& LED & 6.2 hr & \textbf{1$\times$} & 8.8\% \\
& Circuit Breakers & 1.0 hr & \textbf{1$\times$} & 7.3\% \\
& EVA (text) & \textbf{0.4 hr} & \textbf{1$\times$} & \textbf{6.7\%} \\
\midrule
\multirow{4}{*}{InternVL3.5-8B} 
& JailGuard & - & 8$\times$ & 13.1\% \\
& AdaShield-A & 1.2 hr & 1.05$\times$ & 3.0\% \\
& VLGuard & 2.0 hr & \textbf{1$\times$} & 1.8\% \\
& EVA     & \textbf{0.8 hr} & \textbf{1$\times$} & \textbf{0.9\%} \\
\bottomrule
\end{tabular}
}
\end{table}

\subsection{Efficiency Evaluation}
\label{subsec:efficiency}
We evaluate the efficiency of \textit{EVA} considering training time and inference overhead. The experiments are conducted on a single NVIDIA A40 GPU, with results averaged over 5 runs. For text-only tasks, we use Vicuna-7B-v1.5, while for multimodal tasks, we employ InternVL3.5-8B as the base model. Table \ref{tab:efficiency} presents the comparison results.
\textit{EVA} demonstrates superior efficiency across both LLMs and VLMs. For LLMs, it requires only 0.4 hours for training and maintains 1$\times$ inference overhead, outperforming \textit{SafeDecoding} (1.07$\times$ overhead), \textit{LED} (6.2 hours of training), and \textit{Circuit Breakers} (1.0 hour of training).
For VLMs, \textit{EVA} achieves competitive robustness with the lowest ASR while maintaining efficiency. Compared with \textit{VLGuard}, \textit{EVA} uses a substantially smaller amount of training data (200 vs. 2977) yet attains a lower ASR (0.9\% vs. 1.8\%) and a shorter training time (0.8 hr vs. 2.0 hr). Moreover, unlike \textit{JailGuard} and \textit{AdaShield-A}, which introduce inference overheads of 8$\times$ and 1.05$\times$, \textit{EVA} preserves 1$\times$ inference cost, making it more suitable for deployment.


\subsection{Ablation Studies}
\label{subsec:ablation}
\partitle{Number of editing visual tokens}
We further investigated the impact of the number of edited image tokens on defense performance. Using attention scores as the selection criterion, we varied the number of edited tokens from 1 to 3 and evaluated the performance on the InternVL3.5-8B. The results are presented in Table~\ref{tab:num_tokens}.
As shown in the table, editing one image token (Top 1) is sufficient to achieve a substantial reduction in ASR (from 0.48 to 0.03) while even slightly improving the model's general utility (69.3 vs. 69.1). Although increasing the number of edited tokens to 2 or 3 further suppresses the ASR to zero, it leads to a precipitous decline in the model's utility. This indicates that editing only the most critical token offers the optimal trade-off between safety and utility.
\begin{table}[htbp]
\centering
\caption{Comparison of the number of edited image tokens using InternVL3.5-8B. \textbf{Bold} indicates the best result in each column.}
\resizebox{0.8\linewidth}{!}{
\begin{tabular}{lccc}
\toprule
\textbf{Method} & \textbf{MM-SafetyBench}     & \textbf{MM-Vet-v2} \\
                       & ASR $\downarrow$   & Score $\uparrow$ \\
\midrule
Vanilla      & 0.48                    & 69.1 \\
\midrule
Top1 token   & 0.03                    & \textbf{69.3} \\
Top2 tokens  & 0.01                    & 45.4 \\
Top3 tokens  & \textbf{0.00}           & 19.7 \\
\bottomrule
\end{tabular}
}
\label{tab:num_tokens}
\end{table}

\partitle{Harmful text tokens identification sources}
To assess the robustness and practicality of \textit{EVA}, we evaluate three different sources for harmful token identification: GPT-4o (ours), Qwen2.5-72B-Instruct, and Llama2-7B-chat. Table~\ref{tab:token_sources} reports the resulting defense performance and utility.
GPT-4o achieves the lowest ASR and the highest MT-Bench score, indicating that more accurate harmful token identification can simultaneously strengthen safety and better preserve model utility. Qwen2.5-72B-Instruct attains performance close to GPT-4o, with moderately higher ASR but comparable MT-Bench score and over 90\% token overlap with the GPT-4o, suggesting it is a strong open-source alternative. Llama2-7B-chat yields clearly higher ASR and lower MT-Bench score. In practice, the token identification model can be chosen according to deployment budgets and openness requirements. GPT-4o provides the strongest overall performance, while Qwen2.5-72B-Instruct offers a competitive option.
\begin{table}[htbp]
\centering
\caption{Comparison of different harmful-token identification sources. \textbf{Bold} indicates the best result in each column.}
\resizebox{\linewidth}{!}{
\begin{tabular}{lccc}
\toprule
\textbf{Token Source} & \textbf{Token Overlap}   & \textbf{Llama2-GCG} & \textbf{MT-Bench} \\
                      & \textbf{with GPT-4o (\%)}   & Avg. ASR$\downarrow$ & Score$\uparrow$ \\
\midrule
GPT-4o (ours)    & \textbf{100.0} & \textbf{0.0025} & \textbf{6.31} \\
Qwen2.5-72B-Instruct      & 92.3  & 0.0050          & 6.12 \\
Llama2-7B-chat        & 54.5  & 0.0175          & 5.39 \\
\bottomrule
\end{tabular}
}
\label{tab:token_sources}
\end{table}

\partitle{Impact of regularization metrics}
We conduct a study to examine the impact of different regularization metrics within our method: Jensen-Shannon divergence (JS) \cite{lin2002divergence}, Cosine similarity (COS) \cite{salton1975vector}, and KL-divergence \cite{kullback1951information} (ours). The results are summarized in Table~\ref{tab:regularization_metrics}.
\begin{table}[htbp]
\centering
\caption{Comparison of different regularization metrics within \textit{EVA}. Report the Avg. ASR across all attack scenarios. Utility is measured by MT-Bench for Vicuna-7B-v1.5 and MM-Vet-v2 for Qwen2.5-VL-7B. \textbf{Bold} indicates the best result in each column.}
\resizebox{\linewidth}{!}{
\begin{tabular}{l|cc|cc}
\toprule
\multirow{2}{*}{\centering \textbf{Metrics}} & \multicolumn{2}{c|}{\textbf{Vicuna-7B-v1.5}} & \multicolumn{2}{c}{\textbf{Qwen2.5-VL-7B}} \\
 & Avg. ASR $\downarrow$ & MT-Bench $\uparrow$ & Avg. ASR $\downarrow$ & MM-Vet-v2 $\uparrow$ \\
\midrule
\centering Vanilla   & 0.8208          & 6.77          & 0.2283            & 56.9 \\
\midrule
\centering JS        & 0.0733          & 6.63          & 0.0033            & \textbf{55.6} \\
\centering COS       & 0.0683          & 6.44          & 0.0027            & 55.1 \\
\centering KL (Ours) & \textbf{0.0675} & \textbf{6.84} & \textbf{0.0005}   & 55.5 \\
\bottomrule
\end{tabular}
}
\label{tab:regularization_metrics}
\end{table}
KL-divergence consistently offers the best trade-off. On Vicuna-7B-v1.5, KL outperforms other metrics with the lowest ASR (6.75\%) and highest MT-Bench (6.84). Similarly, on Qwen2.5-VL-7B, it reduces ASR to a negligible 0.05\% while maintaining multimodal utility (55.5) comparable to the best-performing baseline. In contrast to JS and COS, which show inconsistent effectiveness across models, KL ensures superior safety without compromising general capabilities.

\partitle{Impact of optimization target ($\mathbf{y}_{safe}$)}
A key choice in \textit{EVA} is the form of the safety target $\mathbf{y}_{safe}$. We compare two strategies: (i) \textit{Universal Refusal}, where all harmful token keys are mapped to a single generic response (e.g., \textit{``I'm sorry, I can't assist with that.''}); and (ii) \textit{Specific Refusal}, where the model is encouraged to generate comprehensive, query-specific refusals that provide detailed justifications and explanations tailored to each harmful query. We conduct experiments on Llama-3.1-8B-Instruct and InternVL3.5-8B. Safety is evaluated by ASR, and utility is evaluated by MT-Bench for the LLMs and MM-Vet-v2 for the VLMs. The results are summarized in Table~\ref{tab:ablation_target}.
\begin{table}[htbp]
\centering
\caption{Ablation on different optimization targets $\mathbf{y}_{safe}$. Universal Refusal vs.\ Specific Refusal on Llama-3.1-8B-Instruct and InternVL3.5-8B. \textbf{Bold} indicates the best result in each column.}
\resizebox{\linewidth}{!}{
\begin{tabular}{l|cc|cc}
\toprule
\multirow{2}{*}{\centering \textbf{Target Type}} & \multicolumn{2}{c|}{\textbf{Llama-3.1-8B-Instruct}} & \multicolumn{2}{c}{\textbf{InternVL3.5-8B}} \\
 & \makecell{GCG \\ Avg. ASR $\downarrow$} & \makecell{MT-Bench \\ Score $\uparrow$} & \makecell{MM-SafetyBench \\ Avg. ASR $\downarrow$} & \makecell{MM-Vet-v2 \\ Score $\uparrow$} \\
\midrule
\centering Vanilla                & 0.5450 & 7.79 & 0.3666 & 69.1 \\
\midrule
\centering Specific               & 0.2125 & \textbf{7.51} & 0.1633 & 68.1 \\
\centering Universal (ours)       & \textbf{0.0075} & 7.44 & \textbf{0.0400} & \textbf{69.3} \\
\bottomrule
\end{tabular}
}
\label{tab:ablation_target}
\end{table}
Across both models, using a \textit{Universal Refusal} target substantially reduces ASR compared to \textit{Specific Refusals}, while MT-Bench and MM-Vet-v2 remain almost unchanged. We see two main reasons for this effect. First, a single universal target yields a consistent optimization signal: all harmful tokens are mapped to the same $\mathbf{y}_{safe}$, so gradients push harmful activations toward one compact refusal region. In contrast, diverse, context-dependent refusals introduce supervision noise and effectively soften the safety boundary, making jailbreaks more likely.
Second, we only adjust a small number of parameters at some layers. With such limited capacity, asking the edit to realize many different, context-specific refusal sentences is hard to optimize and tends not to converge to sufficiently low ASR. A universal target instead concentrates this limited editing capacity on enforcing a single, strong refusal behavior, which empirically produces much stronger safety guarantees. These observations justify our choice of a universal, static $\mathbf{y}_{safe}$ in \textit{EVA}.

%% file: 6_conclusion.tex
\section{Conclusion}
\label{sec:conclusion}
In this work, we propose \textit{EVA}, a safety framework that leverages model editing to robustly defend against jailbreak attacks across both LLMs and VLMs. By precisely updating critical MLP parameters, \textit{EVA} effectively neutralizes harmful behaviors triggered by textual and visual inputs without compromising general model capabilities. Extensive experiments demonstrate that \textit{EVA} achieves best defense performance and utility preservation compared to existing baselines. Overall, this work pioneers the application of direct model editing for VLM safety, offering a unified solution for diverse modalities. Future work could explore extending this mechanism to dynamic inputs, such as Video-LLMs, and developing more lightweight techniques for safety parameter localization.

\section*{Acknowledgments}
We thank the associate editor and anonymous reviewers for their constructive comments and suggestions. This research is supported in part by the “Pioneer” and “Leading Goose” R\&D Program of Zhejiang (Grant No. 2024C01169), the Kunpeng–Ascend Science and Education Innovation Excellence/Incubation Center, the National Natural Science Foundation of China (Grant No. 62441238), and the National Natural Science Foundation of China under Grant U2441240 (“Ye Qisun” Science Foundation).